\documentclass[pra,aps,twocolumn,showpacs]{revtex4}

\usepackage{bm,graphicx,amsmath,amssymb}
\usepackage{bbm}
\usepackage{placeins}
\usepackage{verbatim} 
\usepackage{subfigure} 

\allowdisplaybreaks

\newcommand{\br}[1]{\langle #1|}
\newcommand{\ke}[1]{|#1\rangle}
\newcommand{\bk}[2]{\langle #1|#2\rangle}

\newcommand{\nn}{\nonumber}

\begin{document}

\title{Interfacing microwave qubits and optical photons via spin ensembles }
\author{Susanne Blum,$^1$ Christopher O'Brien,$^{2,3}$ Nikolai Lauk,$^{2}$ Pavel Bushev,$^4$ Michael Fleischhauer,$^{2}$ and Giovanna Morigi$^{1}$}
\affiliation{$^{1}$ Fachrichtung 7.1: Theoretische Physik, Universit\"{a}t des Saarlandes, D-66123 Saarbr\"{u}cken, Germany\\
$^{2}$ Fachbereich Physik und Forschungszentrum OPTIMAS,
Technische Universit\"at Kaiserslautern, D-67663 Kaiserslautern, Germany\\
$^{3}$ Institute for Quantum Science and Engineering, Department of Physics and Astronomy, Texas A \& M University, College Station TX 77843-4242, USA\\
$^{4}$ Fachrichtung 7.2: Experimentalphysik, Universit\"{a}t des Saarlandes, D-66123 Saarbr\"{u}cken, Germany
}

\date{\today}

\begin{abstract}
A protocol is discussed which allows one to realize a transducer for single photons between the optical and the microwave frequency range. 
The transducer is a spin ensemble, where the individual emitters possess both an optical and a magnetic-dipole transition. Reversible frequency conversion is realized by combining optical photon storage, by means of EIT, with the controlled switching of the coupling between the magnetic-dipole transition and a superconducting qubit, which is realized by means of a microwave cavity. The efficiency is quantified by the global fidelity for transferring coherently a qubit excitation between a single optical photon and the superconducting qubit.  We test various strategies and show that the total efficiency is essentially limited by the optical quantum memory: It can exceed 80\% for ensembles of NV centers and approaches 99\% for cold atomic ensembles, assuming state-of-the-art experimental parameters. This protocol allows one to bridge the gap between the optical and the microwave regime so to efficiently combine superconducting and optical components in quantum networks.
\end{abstract}

\maketitle

\section{Introduction}
Hybrid quantum networks combine the long coherence times of microscopic quantum systems with the strong interactions and integration available in solid-state devices \cite{QHybrid}. Their realization requires the efficient interfacing of these constituent systems. In most designs information transport is realized by means of low-loss telecom fibers \cite{Gisin}, while promising candidates for information processing are superconducting quantum circuits, which work in the microwave regime \cite{Clarke2008}. 
One quest is to be able to reversibly convert the frequency of a {\it single} photon from the optical to the microwave regime.

There are several proposals for optical-to-microwave transducers, which make use of different physical mechanisms. One candidate are nanomechanical devices, which strongly couple to both microwave and optical fields via electro- and optomechanical forces, respectively \cite{Stannigel:2010,Vitali:2012,Cleland:2013,Regal:2014,Polzik:2014}. Proof-of-principle experiments  were recently successfully performed \cite{Cleland:2013,Regal:2014,Polzik:2014}. One drawback of nanomechanical systems is that their conversion bandwidth is limited by the high quality factors of the mechanical resonances.

\begin{figure}[ht!]
 \subfigure[]{

 	 \includegraphics[width=4cm]{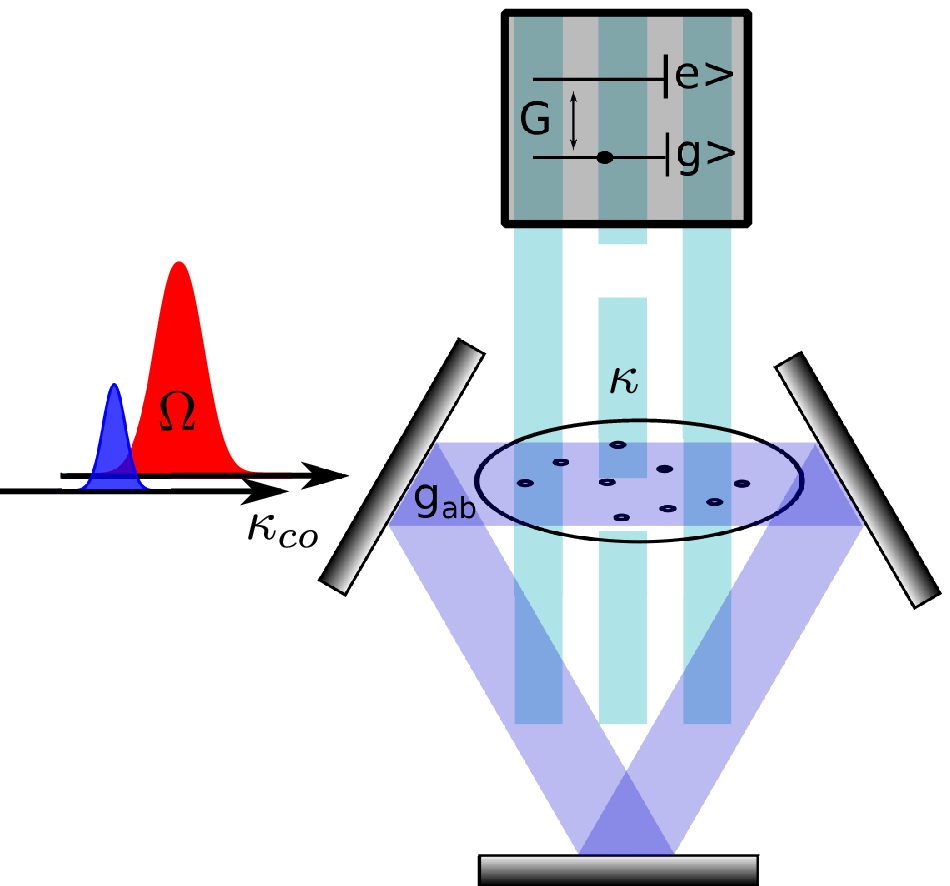}
	 }
\subfigure[]{

 \includegraphics[width=4cm]{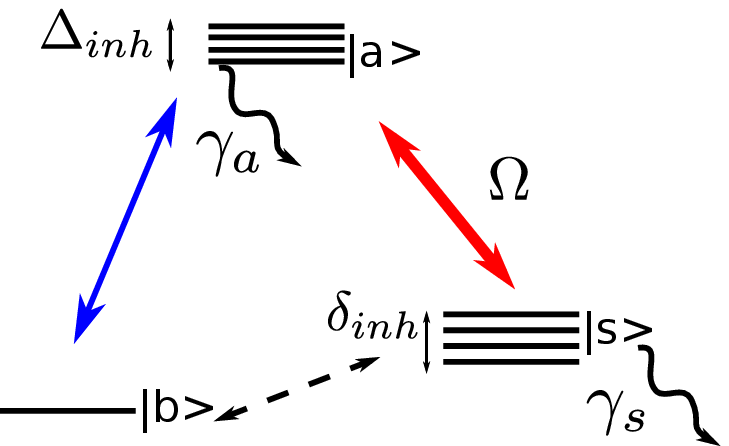}  
}
 \caption{(color online) Left panel: Setup of a quantum transducer based on a spin ensemble. The optical photon (blue) and the control pulse (red) couple with the modes of a ring resonator, which in turn drive the $\Lambda$-level structure of atoms or NV centers in a diamond matrix (Right panel). The photon is stored into a spin excitation along the magnetic-dipole transition $\ke{b}\to\ke{s}$ via EIT. The transition is then coupled to a microwave stripline cavity, which in turn couples to a SC-qubit. The coupling with the microwave resonator is switched on and off either adiabatically or suddenly by tuning the cavity and/or the qubit frequency.  The multiple lines of the level scheme indicate the transitions which are inhomogeneously broadened when the spin ensemble is composed by NV centers. Further details are given in the text. \label{Fig:1}} 
 \end{figure}

An alternative route to conversion could be realized by a spin ensemble, acting as a quantum memory and possessing a magnetic dipole transition which strongly couples to a microwave cavity \cite{Imamoglu:2009}. The basic idea we pursue is the realization of a quantum memory for optical photons. The qubit state is then reversibly transferred from the spin ensemble to a superconducting qubit (SCQ) by controllable switching on and off the coupling of both systems with a microwave resonator, as proposed in Ref. \cite{OBrien:2014}. Examples of spin ensemble are cold atomic gases, Nitrogen-Vacancy (NV) centers in diamonds, and rare-earth doped crystals, of which the strong coupling of a magnetic dipole transition with a microwave resonator has been recently experimentally demonstrated \cite{Verdu:2009, Kubo2011, Staudt2012, Probst2013, Probst2014, Ranjan:2014}. Such a great variety of spin ensembles exhibit different features, which require the application of adequate conversion protocols \cite{Gorschkov:2007}. Rare-earth doped crystals, for instance, are characterized by large optical depths, which makes them an attractive platform \cite{Thiel2011}. Their inhomogeneous broadening,  on the other hand, is such that light storage protocols based on adiabatic transfer, such as storage based on Electromagnetically Induced Transparency (EIT) \cite{Fleischhauer:2005,Lukin2000,Fleischhauer2002}, become inefficient, so that a quantum memory for optical photons shall be realized using methods based on photon echoes \cite{CRIB} and atomic frequency combs \cite{Afzelius:2010}. In Refs. \cite{OBrien:2014,Williamson:2014} it was  shown that control over these techniques can lead to single-photon frequency-conversion efficiencies that exceed 90\%. 
 
In this paper we analyse the efficiency of frequency conversion for other platforms, focusing in particular on cold atomic ensembles \cite{Verdu:2009,Hafezi:2012,Bernon2013} and NV centers  embedded in a diamond matrix \cite{Marcos2010,Zhu2011,Nemoto:2014}. These systems are characterized by limited inhomogeneous broadening, which allows one to implement protocols based on adiabatic transfer. On the other hand, they also possess a low optical depth, so that the setup shall integrate an optical element increasing their coupling with the incident photon, such as an optical resonator.

The protocol we consider in this paper is sketched in Fig. \ref{Fig:1}. The spin ensemble couples with the optical photon via the dipole transition between the electronic states $\ke{b}\to\ke{a}$, and stores the photon into a collective spin excitation via storage based on EIT \cite{Lukin2000,Fleischhauer2002}.  The ring resonator, within which the spin ensemble can be confined, increases the optical depth of the medium.  Coherent transfer of a spin into a qubit excitation is performed by means of a microwave cavity, which couples to both systems and which can be tuned on and off resonance. Our analysis focuses on experimentally accessible parameter regimes and identifies the requirements in order to reversibly transfer photonic to SCQ excitations.

This article is structured as follows. In Sec. \ref{Sec:Model} the physical model is introduced. In Sec. \ref{Sec:Protocol} the strategies considered in this paper for achieving the transfer from optical to microwave are summarized and the efficiency is determined using parameters that are compatible with existing experimental setups. In Sec. \ref{Conclusions} further remarks on the experimental feasibility are made and the conclusions are drawn. The appendices report details on the theoretical model which is at the basis of the results in Section \ref{Sec:Protocol}.


\section{Model}
\label{Sec:Model}

The quantum transducer is a spin ensemble of either NV centers in diamond or cold atoms, which possess an optical transition between the electronic states $\ke{b}$ and $\ke{a}$ coupling with the mode of an optical ring resonator. State $\ke{a}$ also couples via an optical dipole transition to the stable state $\ke{s}$, so that $\ke{b}$, $\ke{a}$, and $\ke{s}$ form a $\Lambda$ configuration of levels, shown in Fig. \ref{Fig:1}(b). The states  $\ke{b}$ and $\ke{s}$ form a magnetic-dipole transition in the microwave regime, which couples with a microwave cavity. 

The protocol is illustrated in Fig. \ref{Fig:1}(a): a single optical photon is coupled to a mode of the ring resonator, and the qubit information it encodes is transferred via light storage into a spin excitation of the transition  $\ke{b}\to\ke{s}$. This spin excitation is then transferred into a SCQ excitation via the coupling with the microwave resonator, which is switched on and off by engineering the cavity and/or the qubit frequency as a function of time. 

In this section we introduce the Hamiltonian governing the coherent dynamics, which is at the basis of the study presented in this work.

\subsection{Hamiltonian}

We first focus on the coherent dynamics of the system, composed by the incident photon, the optical and the microwave cavity modes, and the collective transitions of the atoms, which are driven by external fields. We choose the frame rotating at the carrier frequencies of the driving fields, so that the explicit time dependence in the terms describing the coupling with the lasers is dropped. Details of the transformation from the laboratory frame to this frame are provided in Appendix \ref{App:A}.

The Hamilton operator is conveniently decomposed in the terms giving rise to different elements of the protocol. We first consider a spin ensemble composed by $N$ atoms or NV centers, and denote the energy of the relevant electronic transitions by 
\begin{equation}
\label{H:spin}
\hat H_{\rm spin}=\hbar \sum_i^N \left(\delta_{ab}^{(i)}\hat{\sigma}_{ba}^{(i)\dagger}\hat{\sigma}_{ba}^{(i)}+ \delta_{sb}^{(i)}\hat{\sigma}_{bs}^{(i)\dagger}\hat{\sigma}_{bs}^{(i)}\right)\,,
\end{equation}
where the subscript $i=1,\ldots,N$ labels the particles and $\hat \sigma_{jl}^{(i)}=\ke{j}_i\br{l}$, such that $\hat \sigma_{jl}^{(i)\dagger}\hat\sigma_{jl}^{(i)}=\ke{j}_i\br{j}$ is the projection operator to state $\ke{j}$ of atom $i$ and $\hat \sigma_{jl}^{(i)\dagger}=\hat \sigma_{lj}^{(i)}$. Transition $\ke{s}\to\ke{a}$ is optical, the corresponding raising (lowering) operator is given by $\hat \sigma_{as}^{(i)}=\hat \sigma_{ab}^{(i)}\hat \sigma_{bs}^{(i)}$ ($\hat \sigma_{sa}^{(i)}=\hat \sigma_{as}^{(i)\dagger}$).  For NV centers in diamonds the detunings $\delta_{ab}^{(i)}$ and $\delta_{sb}^{(i)}$ depend on the position $\vec{r}_i$ of the spin $i$ within the sample. 

In the following, we assume that the transition $\ke{b}\to\ke{a}$ couples with the degenerate modes of a ring resonator, while $\ke{s}\to\ke{a}$ is driven by a classical laser pulse. The corresponding Hamiltonian, describing the coupling between spins and optical fields, reads
\begin{eqnarray}
\hat{H}_{\rm opt}
&=&\hbar\sum_{i=1}^N\left[g_{ab}e^{i\vec{k}\cdot \vec{r}_i}\hat{\sigma}_{ba}^{(i)\dagger}\hat{c}_R+{\rm H.c.}\right]\nonumber\\
& & +\hbar \sum_{i=1}^N\left(\Omega(t)e^{i\vec{k}_L\cdot \vec{r}_i}\hat{\sigma}^{(i)\dagger}_{sa}+{\rm H.c.}\right)\nonumber\,.
\end{eqnarray}
Here, $\hat{c}_R$ annihilates a photon of the cavity mode which propagates in clockwise direction, $\vec{k}$ is the corresponding wave vectors, and  $g_{ab}$ is the vacuum Rabi frequency. The laser field drives the transition  $\ke{s}\to\ke{a}$ with spatially-homogeneous strength $\Omega(t)$, which is here assumed to be slowly-varying in time and such that $\Omega(t)=\Omega_0(t){\rm e}^{{\rm i}\phi}$, with $\Omega_0(t)=|\Omega(t)|$. We denote the laser wave vector by $\vec{k}_L$, with $|\vec{k}_L|\approx |\vec{k}|\equiv k$, which is fulfilled considering that the frequency of the transition $\ke{b}\to\ke{s}$ is  in the microwave range. The Hamiltonian term for the mode propagating in the anti-clockwise direction is given in the appendix and is included in the dynamics we evaluate. 

The controlled transfer between the spin excitation and the superconducting qubit occurs via the mode of a microwave resonator at detuning $\Delta_{c}$, which couples with the transition $\ke{b}\to \ke{s}$ of the spins and with the transition $\ke{g}\to \ke{e}$ of the qubit. We introduce the operator $\hat{\sigma}_Q=\ke{g}\br{e}$ and denote by $\delta_Q$ the qubit detuning. The Hamiltonian for this element of the dynamics includes also the quantum field of the microwave resonator and takes the form
\begin{eqnarray}
\label{H:mw}
\hat{H}_{\rm mw}&=&\hbar \delta_Q \hat{\sigma}^\dagger_Q\hat{\sigma}_Q+\hbar \Delta_{c}\hat{a}^\dag\hat{a}\\
& &+\hbar \left[\hat{a}^\dag\left(G\hat\sigma_Q+\sum_{i}^{N}\kappa_i\hat{\sigma}_{bs}^{(i)}\right)+{\rm H.c.}\right]\nn\,,
\end{eqnarray}
where $\hat{a}^\dag$ is the bosonic operator that creates a photon in the microwave cavity mode, while $G$ and $\kappa_i$ denote the vacuum Rabi coupling with the qubit and with the spin at position $\vec{r}_i$, respectively. 

We finally describe the incident photon and the coupling with the resonator. The incident photon is a superposition of the modes external to the resonator and couples with the clockwise cavity mode via a mirror, as is assumed to propagate along the positive direction of the $x$-axis. The corresponding Hamiltonian includes the quantum electromagnetic field outside the resonator, with frequencies $\omega_l$, detuning $\Delta_l=\omega_l-\omega_{co}$, wave vector $k_l$ along the positive direction in the $x$ axis, and corresponding annihilation and creation operators $\hat{d}_l^{(+)}$ and $\hat{d}_l^{(+)\dag}$:
\begin{eqnarray}
\hat H_{\rm in-out}&=&\hbar\sum_l\left[\Delta_l \hat{d}_l^{(+)\dag}\hat{d}_l^{(+)}+ \kappa_{{\rm opt},l}\left(\hat{c}_R^\dag\hat{d}_l^{(+)} +{\rm H.c.}\right)\right]\nn\,,
\end{eqnarray}
where $\kappa_{{\rm opt},l}$ is the coupling parameter between the modes of the optical cavity and the free field, while the mirror where they couple is at the $x=0$ plane. The total Hamiltonian governing the coherent dynamics then reads 
\begin{equation}
\hat H=\hat H_{\rm spin}+\hat{H}_{\rm mw}+\hat{H}_{\rm opt}+\hat H_{\rm in-out}\,.
\end{equation}

\subsection{Target state}

We analyze the dynamics of an incident photon in the Schr\"odinger picture, assuming that the spins are initially ($t=0$) all in state $\ke{b}_i$ and the cavities modes are empty. 
In absence of other sources of excitation, the vector $\ke{\Psi(t)}$ describing the state of the system is composed by the states which contain either one spin or qubit excitation, or a (microwave or optical) cavity photon.  In this regime it is convenient to introduce the following notation for the spin excitations \cite{Fleischhauer2002}: 
\begin{subequations}
\begin{align}
\ke{b}_e& \equiv \ke{b_1,\ldots,b_N} \label{eq:b}\,,\\
\ke{a_i}_e& \equiv \ke{b_1,\ldots,a_i,\ldots,b_N}=\sigma_{ba}^{(i) \dag}\ke{b}_e\label{eq:ai}\,,\\
\ke{s_i}_e& \equiv \ke{b_1,\ldots,s_i,\ldots,b_N}=\sigma_{bs}^{(i) \dag}\ke{b}_e\label{eq:si}\,,
\end{align}
\label{eq:states-atoms}
\end{subequations}
\noindent which compose the set of electronic states involved in the dynamics. Further, we denote by $\ke{vac}$ the vacuum state of the external field, such that one photon excitation in mode  at frequency $\omega_l$ and propagating along the positive direction of the $x$-axis reads $\ke{1_l^{(p)}}=d^{(p)\dagger}_l\ke{vac}$. The cavity-mode states which are relevant to the dynamics are $\ke{0_\ell}$ and $\ke{1_\ell}$, with $\ell=R,L,c\mu$ for the optical clockwise, anti-clockwise, and microwave cavity modes, respectively.  We then conveniently write the state vector at time $t$ in the form:
 \begin{eqnarray}
 \label{eq:psi-red}
 \ke{\Psi(t)}=\ke{\Psi(t)}_{\rm opt}\ke{0_{c\mu}}\ke{g}+\ke{vac}\ke{b}_e \ke{0_L,0_R}\ke{\Psi(t)}_{\rm mw}\,,
 \end{eqnarray}
 where $\ke{\Psi(t)}_{\rm opt}$ is defined in the Hilbert space of external field, optical transitions and cavity modes, and reads:
 \begin{eqnarray}
 \ke{\Psi(t)}_{\rm opt}&=& \sum_l\eta_l(t)|1_l\rangle\ke{0_L,0_R}\ke{b}_e\\
 & &+ \ke{vac}(u(t)\ke{0_L,1_R}+v(t) \ke{1_L,0_R})\ke{b}_e\nn\\
 & &+\sum_i^N\ke{vac}\ke{0_L,0_R}(a_i(t) \ke{a_i}_{e}+s_i(t) \ke{s_i}_{e})\,,\nn
 \end{eqnarray}
 while $\ke{\Psi(t)}_{\rm mw}$ is defined in the Hilbert space of the qubit and of the mode of the microwave cavity:
 \begin{eqnarray}
\ke{\Psi(t)}_{\rm mw}=c(t)\ke{1}_{c\mu} \ke{g}+q(t)\ke{0_{c\mu}}\ke{e}  \,.
 \end{eqnarray}
We remark that, since state $\ke{\Psi(t)}$ is normalized to unity, then $_{\rm opt}\langle\Psi(t)|\Psi(t)\rangle_{\rm opt}+_{\rm mw}\langle\Psi(t)|\Psi(t)\rangle_{\rm mw}=1$. According to this notation, $\eta_l(t)$ is the probability amplitude of a photon in the free-field mode $l$, $u(t)$ and $v(t)$ are the probability amplitudes of a photon in the clockwise and anti-clockwise mode of the ring resonator, $s_i$ and $a_i$ are the probability amplitudes that the ensemble is in state $\ke{s_i}_e$ and $\ke{a_i}_e$ , respectively, $c(t)$ that the excitation is a microwave cavity photon, and finally $q(t)$ that the excitation is transferred in the qubit state $\ke{e}$.

The target is that, given the initial state 
\begin{equation}
\label{state:init}
\ke{\Psi(0)}=\sum_l\eta_l^{(+)}(0)|1_l^{(+)}\rangle\ke{b}_e\ke{0_L,0_R}\ke{0_{c\mu}}\ke{g}\,,
\end{equation}
with $\sum_l|\eta_l^{(+)}(0)|^2=1$, then at a given time $t$ the probability $|q(t)|^2=1$. Ideally, moderate fluctuations in the time $t$ and in the parameters of the protocol shall not significantly affect this probability, which is what we aim for. So far, the dynamics governed by Hamiltonian $\hat H$ does not include detrimental effects, such as decay of the electronic and of the qubit state, dephasing of the internal coherences, and losses of the cavity modes (including also coupling of the other mirrors to the external electromagnetic field). All these effects contribute in decreasing the final probability of transfer, and are here included as loss terms in the equations of motion of the probability amplitudes, which are reported in Appendix \ref{App:equations}.

We observe that our analysis is sufficient to determine the fidelity of transfer of any initial state of the external field of the form
$\ke{\phi_f}=\alpha\ke{\Phi(0)}+\beta\ke{\Psi(0)}$ ($|\alpha|^2+|\beta|^2=1$), where $\langle \Phi(0)|\Psi(0)\rangle=0$ and state $\Phi(0)$ does not couple with the spin ensemble. This holds as long as the transfer process is coherent, as we assume in our model. Losses and decoherence in general lower the probability of transfer of any initial state of this kind into a qubit excitation, and thus the total fidelity.


\section{Protocols for a quantum transducer}
\label{Sec:Protocol}

In this section we discuss strategies for achieving the reversible transfer of a single optical photon into a SCQ excitation. 
In general, the problem can be formulated in terms of optimization, where the various parameters such as detunings and fields are varied as a function of time in order to perform the coherent transfer with unit fidelity. Thus, given the system properties the control parameters are the shape and the intensity of the laser pulse coupling the transition $\ke{a}\to\ke{s}$, the detunings of the individual transitions of ensemble and qubit, and the frequency of the microwave cavity mode \cite{Franca:2005,Koch:2013,Rojan:2014}. While such general analysis has a high computational complexity and deserves a study of its own, here we focus on the efficiency of simple protocols whose components have been proposed in the literature and realized experimentally, which are known to  allow for a certain robustness against parameter fluctuations, and in particular against fluctuations in the time needed to perform the protocol.

There are several cases that are relevant for experimental realizations. Rare-earth-doped crystals exhibit large inhomogeneous broadening and large optical depth. For this case in Ref. \cite{OBrien:2014} we proposed a sequential protocol. There, the incoming photon is absorbed using controlled reversible inhomogeneous broadening (CRIB). The optical excitation is then mapped into a spin state using a series of $\pi$-pulses and subsequently transferred to a superconducting qubit via the microwave cavity. 

In this paper we focus on ensembles of NV centers and of cold atoms. Ensembles of cold atoms in the gas phase \cite{Hafezi:2012} are usually weakly coupled with the external environment, so that the assumption of homogeneous broadening is justified. The atomic motion and the collisional dephasing between the atoms play a minor role during the interaction with the single photon and will be neglected here. We will also assume that the ensemble is small enough so that all atoms feel the same coupling $g_{ab}$ to the optical cavity. In this regime a protocol based on EIT is efficient. NV centers in a diamond crystal act like a frozen gas of atoms, since the centers are embedded in a diamond matrix and thus cannot move. The drawback is that the NV centers are usually inhomogeneously broadened predominantly due to crystal strain and excess nitrogen which is not paired with a neighboring vacancy \cite{Sandner2012}. The latter is actually a limitation to creating high optical depth for the EIT-storage, because higher dopant densities lead to a higher number of unpaired nitrogen atoms and thus to shorter pure optical dephasing times ($T_2^*$) \cite{Nobauer2013}. A way to get rid of this inhomogeneous broadening was proposed in Ref. \cite{Bensky2012} by preselecting the optimal spectral portion of NV centers and transfering the rest to an auxiliary state. This leads to a decrease in the spectral width. At the same time it also leads to a decrease of about $5 \times 10^3$ in the number of spins, which can be partially compensated by means of the amplified coupling induced by a resonator, as we consider here.

Under these premises, we review in this section the basic steps of an EIT-based storage protocol, followed by an adiabatic transfer of the spin excitation to the qubit by slowly switching the microwave cavity. The transfer is sequential and aims at maximizing the fidelity of the individual steps. For a sequential protocol, the total fidelity can be then written as the product of the fidelities of the individual steps of the sequence:
\begin{align}
 \mathcal{F}=\mathcal{F}_{\text{EIT}}\mathcal{F}_{\text{mw}},
\end{align}
where $\mathcal{F}_{\text{EIT}}$ is the fidelity of the light storage protocol, and according to our model it corresponds to the probability of mapping the incident photon into a
spin excitation, while $\mathcal{F}_{\text{mw}}$ is the probability to transfer the collective spin excitation to the qubit.

\subsection{Light storage}

Let us first consider how to map a single photonic excitation onto a spin excitation of a medium. For this purpose we assume that the spin ensemble is not coupled to the microwave cavity field, which can be achieved by setting the cavity mode and qubit out of resonance. The storage protocol we suggest to use is based on EIT \cite{Fleischhauer:2005}, a phenomenon which occurs in a medium consisting of three level atoms in a $\Lambda$ configuration as in Fig. \ref{Fig:1}(b). Applying a resonant control field on the $\ke{s}-\ke{a}$ transition opens a transparency window for the propagating signal field which travels
trough the medium with a reduced group velocity. Since the group velocity depends on the control-field Rabi frequency $\Omega(t)$, this propagating field can be stored by adiabatically reducing the control field strength to zero \cite{Lukin2000}.
A detailed analysis \cite{Fleischhauer2002, Gorshkov:2007:2} shows that the storage efficiency reaches unity in the limit of infinite optical depth $d$. Unfortunately, both gas and NV center ensembles typically have small optical depths.
To circumvent this issue one can use an optical cavity as we consider here, which increases the optical depth by the number of passes which a photon makes through the cavity before leaking out. This hence leads to an effective
optical depth which is then described by the cooperativity parameter $C=4g_{ab}^2N/\gamma_{co}\gamma_a$ of the cavity, where $\gamma_{co}$ and $\gamma_a$ are the cavity and atomic decay rates, respectively and $g_{ab}\sqrt{N}$ is the coupling strength
between cavity mode and atomic ensemble. The maximal efficiency in that case is then given by $\eta_{\text{EIT}}=\frac{C}{1+C}$ \cite{Gorshkov:2007:1}. 
Furthermore, assuming a homogeneously broadened medium and neglecting
decay of the metastable state it was shown in Refs. \cite{Fleischhauer2000b, Gorshkov:2007:1} that in the ``bad cavity''
limit ($\gamma_{co}\gg g_{ab}\sqrt{N}$) any smooth input mode with the duration $T$ which satisfies the adiabaticity condition $TC\gamma_a\gg1$ can be stored with maximally possible efficiency into the target intermediate state
\begin{align}
 \ke{\Psi}_{opt|target}=\ke{vac}\ke{0_L,0_R}\ke{s}_e
\end{align}
by suitably shaping the control-field pulse. Details are reported in Appendix \ref{App:EIT}.
The state $\ke{s}_e=\sum_je^{i(\vec{k}-\vec{k_L})\cdot\vec{r}_j}\ke{s}_j/\sqrt{N}$ describes the stored spin state. If the propagation direction of the control laser coincide with
the one of the incident photon, i.e. $\vec{k_L}\approx\vec{k}$, the phase term can be neglected and the spin state is then approximated by the symmetric Dicke state $\ke{s}_e\approx\sum_j\ke{s}_j/\sqrt{N}$.
The discussion above implies that for optimal transfer the time scale of the storage process should be faster than detrimental effects (which we neglected so far). Among these processes is the dephasing due to inhomogeneous broadening
which is in particular relevant for the NV center ensembles. 

Another source of losses is the coupling with the other ring cavity mode propagating in the counter clockwise direction which is fully included in our analysis. The rate of losses associated with this effect is estimated to be $\lesssim |\sum_j e^{2i\vec{k}\cdot\vec{r}_j}|^2g_{ab}^2/N$, as shown in Appendix \ref{App:EIT}, and can be rather small provided a sufficiently large number of atoms homogeneously distributed
in space over several wavelengths of light.

\subsection{Transfer from a spin to a qubit excitation via a microwave cavity}

We first consider the second part of the protocol, in which the spin excitation is transferred into a qubit excitation via coupling to the microwave resonator. This coupling can be neglected during the light storage protocol assuming the ability to tune the cavity mode and the qubit transition far-off resonance. After storage, the coupling can be switched on in a controlled way by setting these two systems close to resonance with the spin transitions. Therefore, we assume that there is an instant of time $t_1>0$ at which the photon has been absorbed. In particular, $\Omega(t_1)=0$, and the system is in state $\ke{\Psi(t_1)}=\ke{\Psi_{\rm opt}(t_1)}\ke{0_{c\mu }}\ke{g}$, with $\ke{\Psi_{\rm opt}(t_1)}=\ke{vac}\ke{0_L,0_R}\ke{s}_e$. The dynamics at times $t>t_1$ are governed by the sum of Hamiltonians \eqref{H:spin} and \eqref{H:mw}:
\begin{eqnarray}
\label{H:mw:2}
H_{\rm mw}'(t)&=&\hbar \sum_i^N\delta_{sb}^{(i)}\hat{\sigma}_{bs}^{(i)\dagger}\hat{\sigma}_{bs}^{(i)}+\hbar \delta_Q \hat{\sigma}^\dagger_Q\hat{\sigma}_Q+\hbar \Delta_{c }\hat{a}^\dag\hat{a}\nn \\ 
&+&\hbar \left[\hat{a}^\dag\left(G\hat\sigma_Q+\sum_{i}^{N}\kappa_i\hat{\sigma}_{bs}^{(i)}\right)+{\rm H.c.}\right],
\end{eqnarray}
where the qubit (or equivalently: spin) and cavity frequency can be tuned in time: $\delta_Q=\delta_Q(t)$ and $\Delta_c=\Delta_c(t)$, while the coupling with the resonator is constant. This time dependence shall be varied so to coherently transfer the population at time $t_2>t_1$ to the SCQ. From the form of Hamiltonian \eqref{H:mw:2} it is immediately visible that the transfer efficiency is limited by the overlap between the symmetric Dicke state $\ke{s}_e$ obtained by light storage and the collective state $\ke{s'}_e$ maximally coupling with the cavity, $\ke{s'}_e=\sum_j\kappa_j\ke{s_j}_e/\sqrt{\sum_j|\kappa_j|^2}$. The amplitudes $\kappa_j$ are determined by  the cavity mode function, and thus by the structure of a coplanar waveguide resonator mode \cite{Schuster:2010, Verdu:2009}. The fidelity of this transfer process has the upper bound $\mathcal F_{\rm mw}\le F_{\rm max}^S$ given by the overlap $$F_{\rm max}^S=|_e\bk{s}{s'
 }_e|^2\,.$$ 
For the general purpose of this discussion, we will assume $\kappa_j=\kappa$, so that $\ke{s'}_e=\ke{s}_e$. Our aim is to identify the temporal variation of $\delta_Q$ and $\Delta_c$ for which the following equations of motion, 
\begin{subequations}
\label{eq:eom-general}
\begin{align}
\dot{s}(t)&= -i\kappa \sqrt{N}c(t)\label{eq:eom-general_sk}\,,\\
\dot{c}(t)&=-\left(i\Delta_{c}(t)+\frac{\gamma_{c\mu}}{2}\right) c(t)-i\kappa \sqrt{N}s(t)-i G q(t)\label{eq:eom-general_c}\,,\\
\dot{q}(t)&= -\left(i \delta_{Q}(t)  +\frac{\gamma_e}{2}\right) q(t)  -i G c(t)\label{eq:eom-general_q}\,,
\end{align}
\end{subequations}
couple the initial state $s(t_1)=1$ with the target state at $t_2>t_1$ with $q(t_2)=1$. In particular, 
$$ \mathcal F_{\rm mw}=|q(t_2)|^2\,,$$
under the condition that $s(t_1)=1$, unless otherwise stated. In this process we search for solutions which are robust against time and parameter fluctuations, and still sufficiently fast to minimize the detrimental effects, such as for instance qubit and cavity decay. We will focus on two different schemes based on (i) adiabatic transfer of the excitation by tuning the qubit or the cavity frequency \cite{Kubo2011}, and on (ii) pulsed transfer by tuning the cavity field so to perform a dynamics which realizes an effective $\pi$-pulse. For convenience, from now on we will restrict our analysis to the states of the restricted Hilbert space composed by the microwave cavity, the spin, and the qubit states. 

\subsubsection{Adiabatic transfer}

\begin{figure}[ht!]
\subfigure[]
   {
\includegraphics[width=7cm]{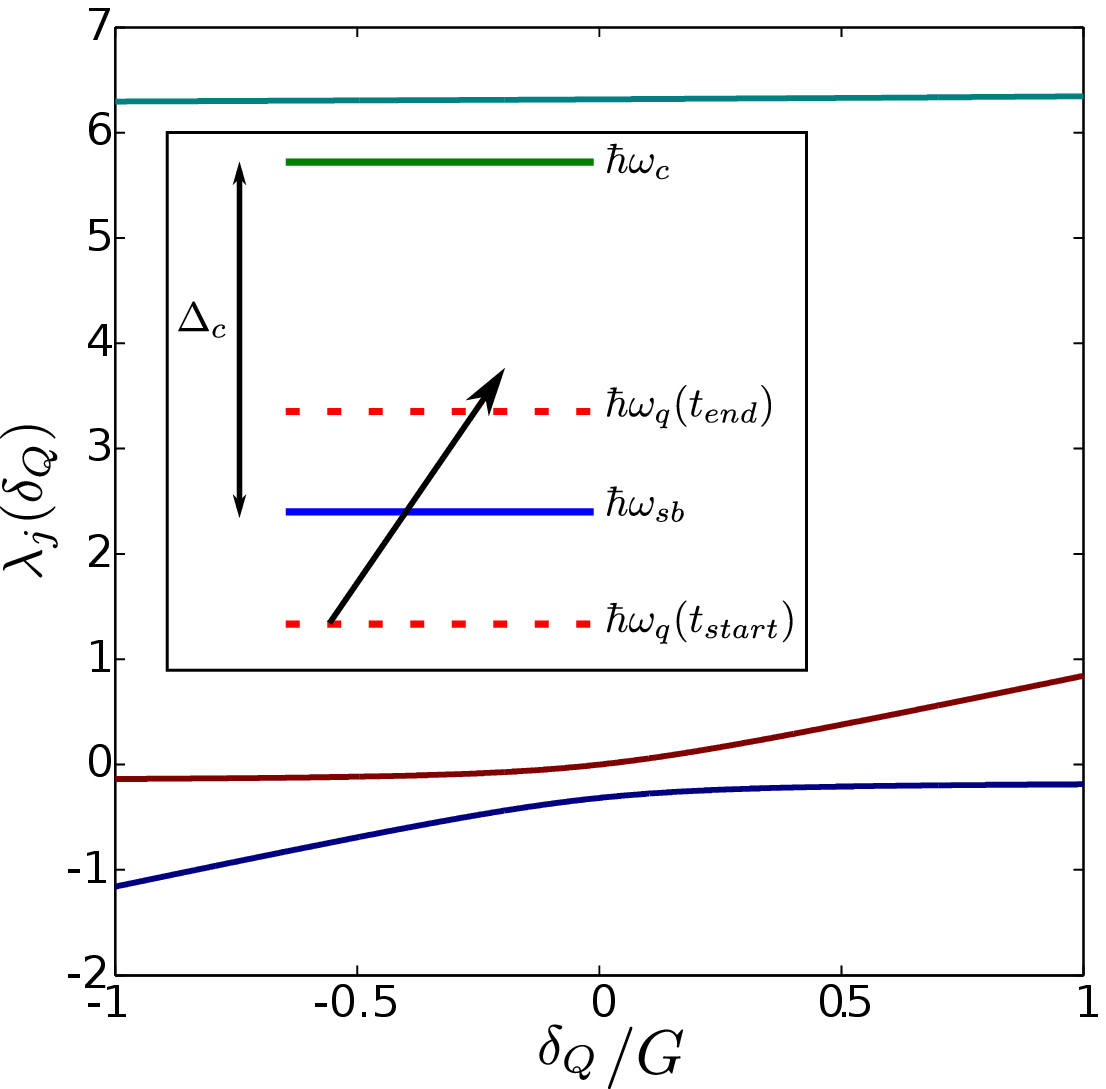}}\label{fig:2a}
\subfigure[]{
\includegraphics[width=7cm]{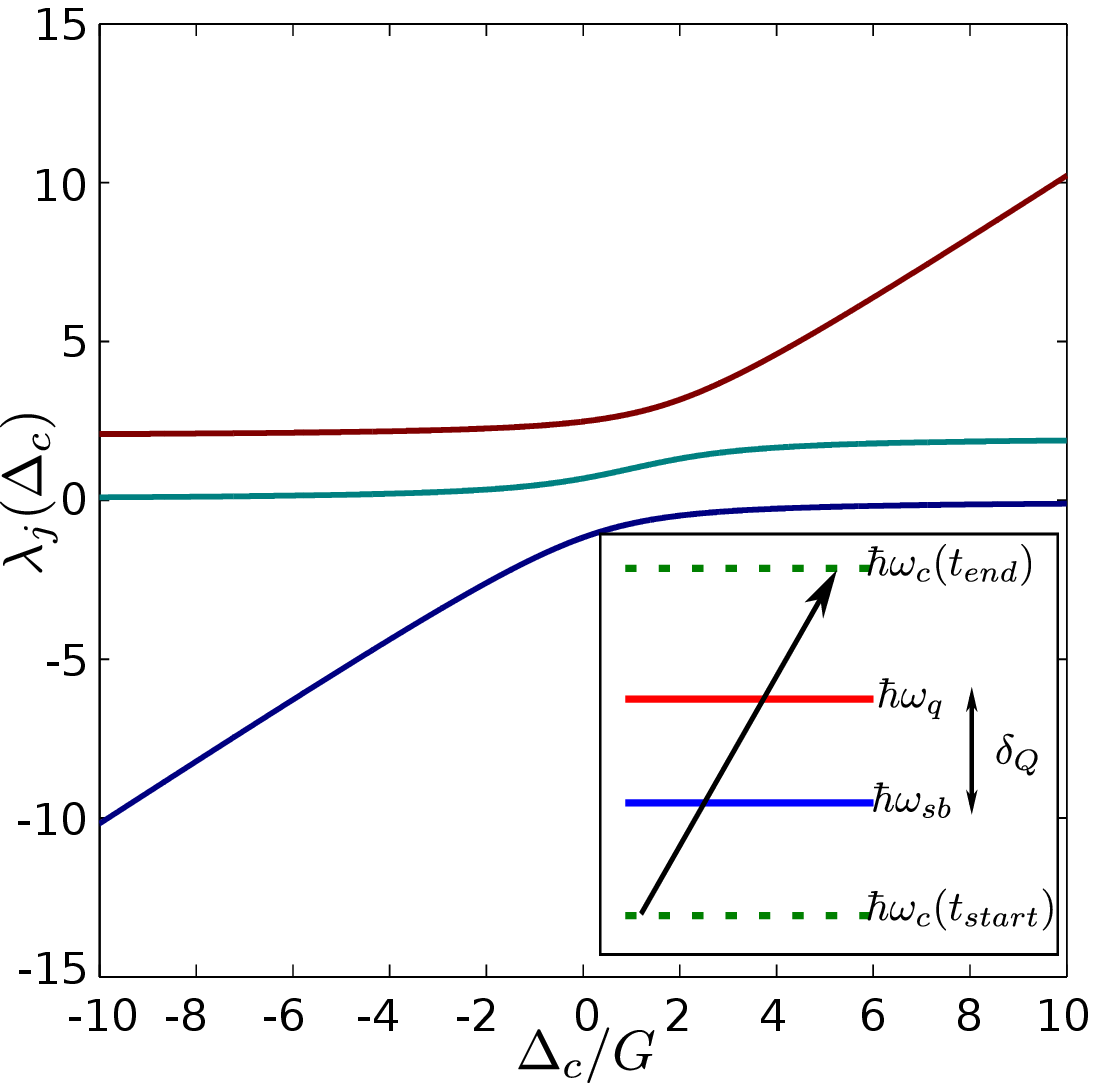}\label{fig:2b}}
 \caption{(color online)  Instantaneous eigenvalues of the three level system as a function of $\delta_Q(t)$ (a) and of $\Delta_c(t)$ (b). The duration of the adiabatic protocol is determined by the energy gap at the avoided crossings. The insets illustrate the time variation of the qubit (a) and of the cavity detuning (b). In (b) the cavity frequency is swept first across the spin and then across the qubit resonance.  \label{Fig:2tune}}
\end{figure}  

We now focus on adiabatic transfer mechanisms, which are realized by changing the cavity and/or the qubit frequency sufficiently slow, so that the system follows the instantaneous eigenstates of Hamiltonian $H_{\rm mw}'(t)$, Eq. \eqref{H:mw:2}, which we denote by $\ke{\lambda_j(t)}$ such that $H_{\rm mw}'(t)\ke{\lambda_j(t)}=\lambda_j(t)\ke{\lambda_j(t)}$, with $j=1,2,3$. The adiabatic transfer corresponds to move along one of the paths shown in Fig. \ref{Fig:2tune} by varying a transition frequency in time \cite{Fewell1997}: when the initial state is an eigenvector at $t\to -\infty$,  this is continuously transformed into the target state, which is an eigenvector at the other asymptotic, $t\to\infty$. The rate of change, which warrants the adiabatic following, must be smaller than the minimal energy gap between the eigenvalues. Thus, if $\ke{\Psi(t_1)}=\ke{\lambda_1(t_1)}$, with $\ke{\Psi(t_1)}=\ke{s}_e\ke{0_{c\mu}}\ke{g}$, and $\ke{\lambda_1(t_2)}=\ke{b}_e\ke{0_{c\mu}}\ke{e}$, then the adiabatic transfer is warranted provided that the condition below is fulfilled at any instant of time $t$ \cite{Schiff}: 
\begin{equation}
\label{eq:adiabcond0}
 |\bk{\dot{\lambda}_1(t)}{\lambda_j(t)}|\ll |\lambda_1(t)-\lambda_j(t)|\,,
 \end{equation}
for $j=2,3$. 

One simple limiting case is found by adiabatically tuning the qubit frequency while keeping the cavity mode far-off resonance, as illustrated in the inset of Fig. \ref{Fig:2tune}(a). In this way, the cavity field is only virtually excited and the effect of cavity decay on the protocol fidelity is minimized. This can be realized when $|\Delta_c|\gg G,\kappa\sqrt{N},\max_t|\delta_Q(t)|$: then the dynamics can be reduced to an effective two-level system, where qubit and spin excitations are directly coupled with rate $\tilde\kappa =G\kappa\sqrt{N}/\Delta_c$, and \begin{align}
 \label{eq:adiabelimeval}
 \lambda_{1,2}(t)&=\frac{\kappa^2N}{\Delta_c}+\frac{\tilde\delta_Q(t)}{2}\mp\sqrt{\frac{\tilde\delta_Q(t)^2}{4}+\tilde{\kappa}^2}\,.
 \end{align}
Here, $\tilde{\delta}_Q(t)=\delta_Q(t)+G^2/\Delta_c-\kappa^2N/\Delta_c$ includes the dynamical Stark shift. The corresponding eigenvector takes the form $\ke{\lambda_1(t)}=\cos\Theta(t)\ke{s}_e\ke{g}+\sin\Theta(t)\ke{b}_e\ke{e}$, where  $ \tan(\Theta)=\tilde\kappa/\lambda_1(t)$, and $\delta_Q$ varies from a negative to a positive value as a function of time. Here, we assume it follows the linear relation $\delta_Q(t)=\delta_{qk}/2(2t/T-1)$ with $t\in [0,T]$ and $\delta_Q^{(0)}\gg|\tilde\kappa|$. The duration $T$ must fulfill the condition dictated by Eq. \eqref{eq:adiabcond0}, thus $|\tilde\kappa| T\gg 1$.  This condition can be optimized choosing different types of time sweeps, which are faster where the energy gap is larger. 

We first analyze with equations \eqref{eq:eom-general} the fidelity of the protocol based on sweeping the frequency of the SCQ through the resonance frequency of the symmetric spin state, while keeping the cavity off resonance, as illustrated in Fig. \ref{Fig:2tune}(a). This requires to sweep the detuning $\delta_Q$ sufficiently slow across resonance, so that the system remains in the state whose energy is given by the upper red line of Fig. \ref{Fig:2tune}(a) and the spin excitation is ideally adiabatically transformed into the qubit excitation at the end of the sweep. In Ref. \cite{Hofheinz2008}, the frequency of a phase qubit can be tuned between 6 and 9\, GHz, and we here assume a tuning range $\delta_{qk}= 3$ GHz. The duration of the protocol $T$ is chosen to be shorter than 10$\mu s$, which corresponds to the lifetime of the SCQ  \cite{Chow2012}. We further take the qubit-cavity coupling to be $G/2\pi=50 $ MHz  \cite{Sillanpaa2007, Blais2004} and set it, for later convenience, equal to  the superradiant coupling of the cavity with the spins, $\kappa\sqrt{N}=G$. 

\begin{figure}[ht!]
\includegraphics[width=7cm]{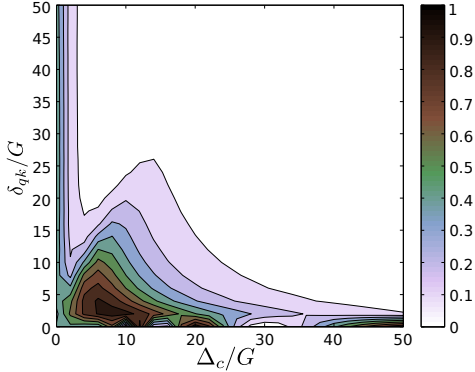}
\caption{(color online) Fidelity $\mathcal F_{\rm mw}$ of transferring the spin into the SCQ excitation for the protocol based on sweeping the qubit detuning across the spin resonance. The fidelity is reported as a function of the cavity detuning $\Delta_{c}$ and of the range $\delta_{qk}$ over which the qubit detuning is linearly swept and is obtained by numerically integrating Eqs. \eqref{eq:eom-general}. The transfer time is $T=100G^{-1}$ and $G=\sqrt{N}\kappa=2\pi \times 50$ MHz. }
\label{fig:qucontour100}
\end{figure}

\begin{figure}[ht!]
    \subfigure[]{
    \includegraphics[width=7cm]{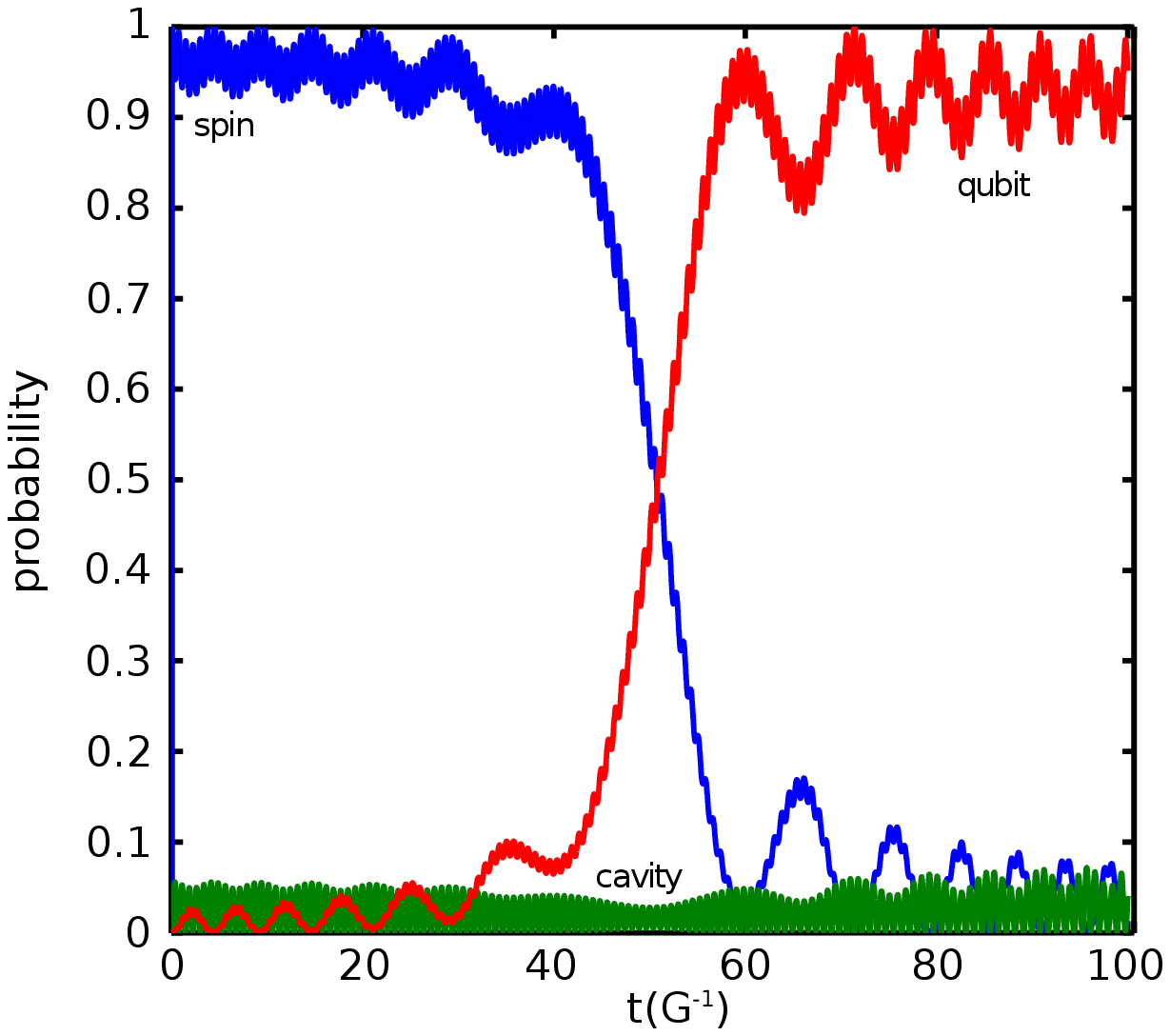}\label{fig:qub0_2ck8qk3duration100}}
\subfigure[]{
 \includegraphics[width=7cm]{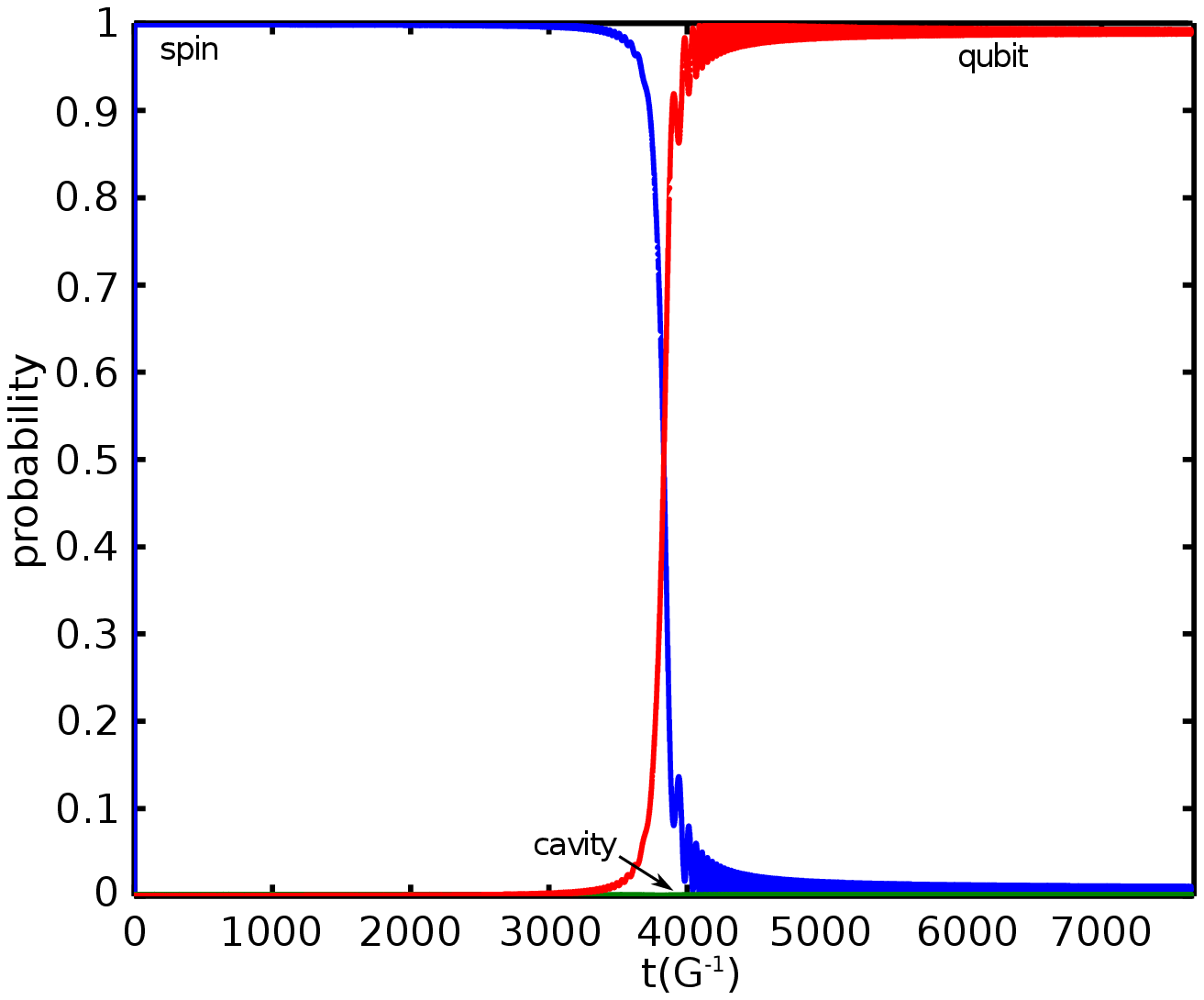}
 \label{fig:qub1ck50qkt4d7650dt0_1}
}
\caption{(color online) Dynamics of the spin (blue), cavity (green), and qubit (red line) populations as a function of time (in units of $G$) during the linear sweep of the qubit detuning as $\delta_Q(t)=\delta_{qk}/2(2t/T-1)$ for $G=\sqrt{N}\kappa=2\pi \times 50$ MHz. The other parameters are (a) $\delta_{qk}=3G$, $\Delta_{ck}=8G$ for a transfer time $T=100G^{-1}$. Here the averaged qubit population at the end of the sweep is approximately 94\%. In (b) $\delta_{qk}=4G$ and $\Delta_{ck}=50G$ for $T=7650G^{-1}$. The average qubit population at the end of the sweep is 99.05\%.}
 \label{Fig:tunequbit}
 \end{figure}

Figure \ref{fig:qucontour100} displays the fidelity $\mathcal F_{\rm mw}$ as a function of the cavity detuning $\Delta_c$ and of the range $\delta_{qk}$ over which the qubit detuning  is swept. Note that, since the transfer time is here kept constant, larger values of $\delta_{qk}$ imply faster sweeps, so there is an optimal parameter regime for which the protocol is efficient. The parameter region with best fidelity corresponds to finite values of $\Delta_c$ (namely, namely finite gaps) and sufficiently small values of $\delta_{qk}$: in this regime, in fact, one has slow quench rates and sufficiently large tuning ranges of the qubit detuning, so to allow one to performing adiabatic sweeping. We observe that average transfer fidelities over 0.9 can be reached in a sufficiently broad region of parameter centered at $\delta_{qk}\sim 3G$ and $\Delta_c\sim 8 G$. The values for the fidelity in Fig. \ref{fig:qucontour100} correspond to the occupation of the SCQ excited state at the end of the sweep, which is obtained by averaging over the oscillations observed at the asymptotics. These oscillations are visible in the qubit population shown in Fig \ref{Fig:tunequbit}(a), which displays the dynamics of spin, cavity, and qubit excitation as a function of time, while the detuning is swept across the spin resonance. They are due to the fact that, for the chosen values of $\delta_{qk}$, the protocol does not start sufficiently far away from the avoided crossing, so that in the final state both cavity and qubit are populated. Suppression of these oscillations, leading to larger fidelities, can be reached for larger values of $\delta_{qk}$ and/or of the detuning $\Delta_c$, provided that the transfer times is scaled up, as shown in Fig. \ref{Fig:tunequbit}(b). Here, we took $\delta_{qk}=4G$ and $\Delta_c=50G$. The final occupation is about 99\%, but the time required for the transfer is one order of magnitude larger, $T=7650 G^{-1}$ and the system suffers from spontaneous decay. 

\begin{figure}[ht!]
\includegraphics[width=7cm]{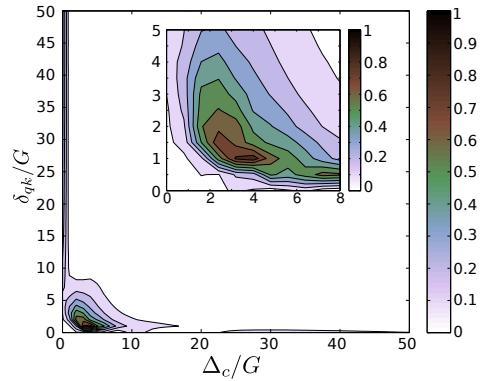}
\caption{(color online) Same as in Fig. \ref{fig:qucontour100} but for $\sqrt{N}\kappa=0.2G$.}
\label{fig:qucontour100b0_2}
\end{figure}

The fidelity of this protocol is sensitive to imbalances in the values of the coupling constants, and requires that the coupling strength $G$ and $\kappa\sqrt{N}$ are of the same order of magnitude. An example of the transfer fidelity  for $\kappa\sqrt{N}=0.2G$ is reported in Fig. \ref{fig:qucontour100b0_2} and shows a severe reduction of the parameter region where the protocol is efficient.  

\begin{figure}[ht!]
\includegraphics[width=7cm]{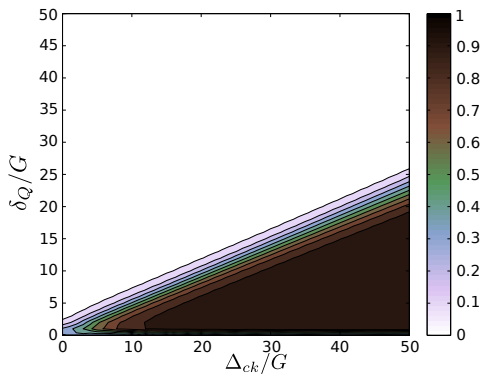}
\caption{(color online) Fidelity of transferring the spin into the qubit excitation for the protocol based on sequentially sweeping the cavity detuning across the spin and then the qubit resonance. The fidelity is reported as a function of the qubit detuning $\delta_{Q}$ and of the frequency range $\Delta_{ck}$ over which the cavity detuning is linearly swept and is obtained by numerically integrating Eqs. \eqref{eq:eom-general}. The transfer time is $T=100G^{-1}$ and $G=\sqrt{N}\kappa=2\pi \times 50$ MHz.}
\label{fig:cavcontour100}
\end{figure}

\begin{figure}[ht!]
 	 \includegraphics[width=7cm]{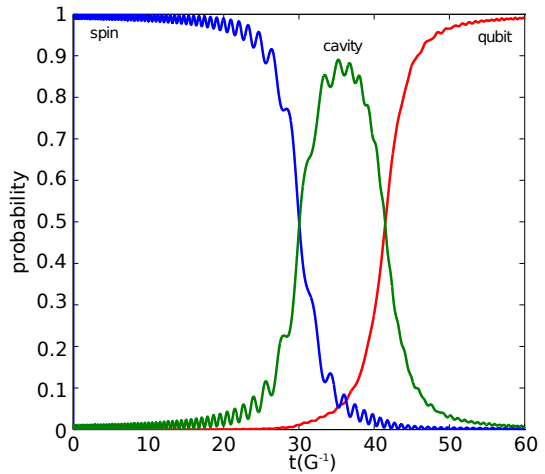}
\caption{(color online) Dynamics of the spin (blue), cavity (green), and qubit (red line) populations as a function of time (in units of $G$) during the linear sweep of the cavity detuning as $\Delta_c(t)=\Delta_{ck}/2(2t/T-1)$ for $G=\sqrt{N}\kappa=2\pi \times 50$ MHz, $\delta_{Q}=7.4G$ and $\Delta_{ck}=38G$. At the time $t=60G^{-1}$ the SCQ excited-state population is slightly above 0.99.}
\label{fig:cavfin}
 \end{figure}

Another simple type of adiabatic protocol is depicted in Fig. \ref{Fig:2tune}(b). It sequentially transfers population (i) from spin to cavity, by keeping the qubit far-off resonance  and sweeping the cavity frequency across the spin resonance, and then (ii) from cavity to qubit, by sweeping the cavity frequency through the qubit resonance. It thus combines a sweep of the cavity detuning $\Delta_c$ and/or of the qubit detuning $\delta_Q$. This sequential transfer requires the excitation of the microwave resonator and is thus sensitive to cavity losses. In order to preserve adiabaticity, the transfer time $T$ must  be larger than $1/\kappa\sqrt{N}$ for the first part and than $1/G$ for the second part. Figure \ref{fig:cavcontour100} displays the fidelity of the protocol, performed by sweeping $\Delta_c$ while keeping $\delta_Q$ and $T$ constant. This corresponds to ideally  following the middle curve in fig. \ref{Fig:2tune}(b) in the limit of sufficiently large $\Delta_{ck}$, under the assumption that the cavity frequency sweeps through the qubit frequency as shown in the inset of fig. \ref{Fig:2tune}(b). We expect then that higher fidelities are found for large $\Delta_{ck}$ and small $\delta_Q$. The results we found in figure \ref{fig:cavcontour100} agree with this expectation. As in the previous case, we set $T=100G^{-1}$ with $G=\kappa\sqrt{N}=2\pi \times 50$ MHz. The detuning $\Delta_c$ is swept linearly with time, and the maximum range is denoted by $\Delta_{ck}$. The region of parameters at fidelity above 0.99 is manifestly larger. Figure \ref{fig:cavfin} displays the time evolution of the occupation probability for a particular choice of the parameters warranting a fidelity of 0.99.

\begin{figure}[ht!]
\includegraphics[width=7cm]{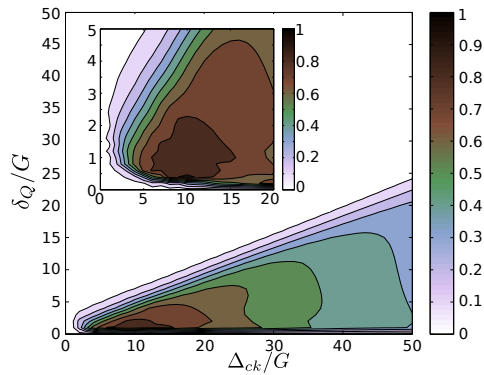}
\caption{(color online) Same as in Fig. \ref{fig:cavcontour100} but for $\sqrt{N}\kappa=0.2G$.}
\label{fig:cavcontour100b0_2g}
\end{figure}

Figure \ref{fig:cavcontour100b0_2g} displays the fidelity of the protocol for the same transfer time as in Fig. \ref{fig:cavcontour100} but when the strength of the collective coupling of spins with the resonator is reduced by a factor 5, namely, for $\sqrt{N}\kappa=0.2G$. The parameter regions of fidelities above 0.9 is drastically reduced, due to the fact that the gap at the avoided crossing (\ref{Fig:2tune}(b)) is reduced and thus strict adiabaticity would now require longer transfer times. Nevertheless, this parameter region is still substantially larger than the one found for the protocol, where the qubit detuning is sweeped across resonance (compare with Fig. \ref{fig:qucontour100b0_2}). 

We note that the tuning range of a  superconducting cavity is to a maximum of 1GHz \cite{Sandberg2008, Osborn2007} for the coupling strength $G$ we considered here. This protocol seems thus more efficient than the one based on sweeping the qubit frequency keeping the cavity far-off resonance. Its drawback is that the probability that one photon is in the cavity during the protocol is appreciable, hence cavity decay (which was not accounted for in this calculation) can affect its efficiency.  Before concluding this section, we note that we just focused on linear sweeps. Further optimization can be performed in shaping the time variation of the detuning, which for the case of the cavity sweep can lead to better fidelities for the same transfer times.

\subsubsection{$\pi$-pulse transfer}
\label{Subsubsec:Mod:Trans:pi}

We now turn to a protocol based on setting spin, qubit, and cavity frequencies on resonance for a time duration corresponding to an effective $\pi$ pulse, which ideally transfers the population from the spin to the qubit. This means that over an interval of time to be identified one has $\delta_{sb}=\delta_Q=\Delta_c=0$. With this assumptions Eqs. \eqref{eq:eom-general}  can be simply solved analytically, and after assuming $s(0)=1$, $c(0)=q(0)=0$ ($t_1=0$), one finds that the probability amplitude that the qubit is in the excited state at $t>0$ is
\begin{equation}
\label{eq:simplesolq}
q(t)= -\frac{2\kappa \sqrt{N}G}{{\Omega}^2}\sin^2\left(\frac{\Omega t}{2}\right)\,,\\
\end{equation}
where ${\Omega}=\sqrt{G^2+\kappa^2N}$. The frequency $\Omega$ thus determines the duration of the transfer pulse, which here means that after the time interval $T=\pi/\Omega$ the qubit and the cavity field are set off resonance. This condition (as well as Eq. \eqref{eq:simplesolq}) is specifically valid for a rapid change of the detunings, so that in a very short time they are set on or off resonance, and is clearly modified if one assumes a smooth time variation. 

\begin{figure}[ht!]
 	 \includegraphics[width=7cm]{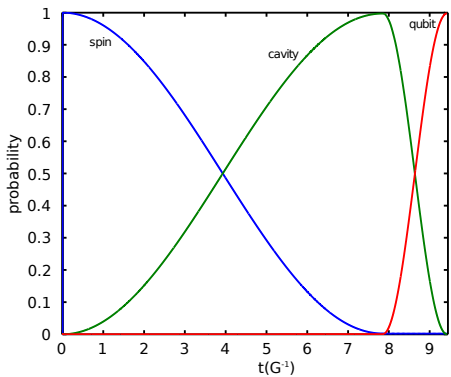}
    \caption{(color online) Dynamics of the populations as a function of time during the protocol, realizing sequential $\pi$ pulses from spin to cavity, and then from cavity to qubit. The parameters are  $\sqrt{N}\kappa=0.2G$, $\delta_{Q}=50G$ and $\Delta_{c}=0$ in the first part of the sequence, then $\Delta_{c}=50G$ in the second part.  \label{fig:doublepi0_2}}
 \end{figure}

Equation \eqref{eq:simplesolq} shows that perfect transfer to the qubit excitation occurs provided that the couplings are matched, $G=\kappa \sqrt{N}$. Mismatching will provide another boundary to the maximal fidelity. For instance let $\epsilon=|G-\kappa\sqrt{N}|/\Omega$ be a relative measure of the mismatch, then the maximal qubit excitation will be $\max_t|q(t)|^2=(1-\epsilon^2)^2$. The final fidelity of the protocol will be thus limited by the upper value  $\mathcal F_{\rm mw}\le F_{\rm max}^S(1-\epsilon^2)^2$. This problem can be solved by performing two subsequent $\pi$-pulses: the first transfers the excitation to the cavity, while the qubit is decoupled, the second from the cavity to the qubit. In this case the time required is $\frac{\pi}{2}(1/G+1/\kappa\sqrt{N})$ (for square pulses ). Taking $G=2\pi\times 50$MHz and $\sqrt{N}\kappa=0.2$G, ideally perfect transfer is achieved over a time of the order of $T\sim10G$. The corresponding dynamics is shown in Fig. \ref{fig:doublepi0_2} and achieves fidelities above 0.999 for $T<10G$.  

\begin{figure}[ht!]
\subfigure[]
   {
\includegraphics[width=7cm]{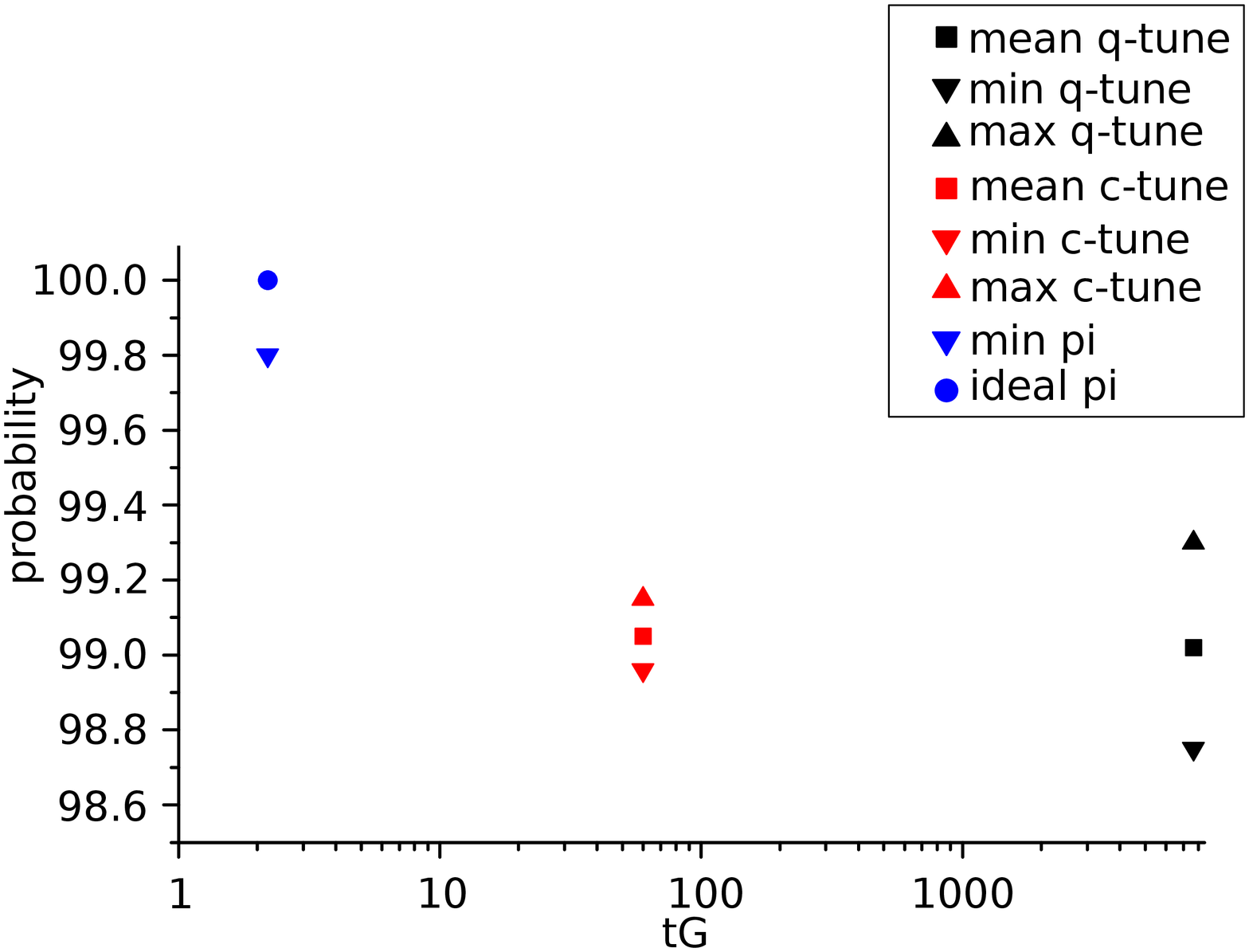}}
\subfigure[]{
\includegraphics[width=7cm]{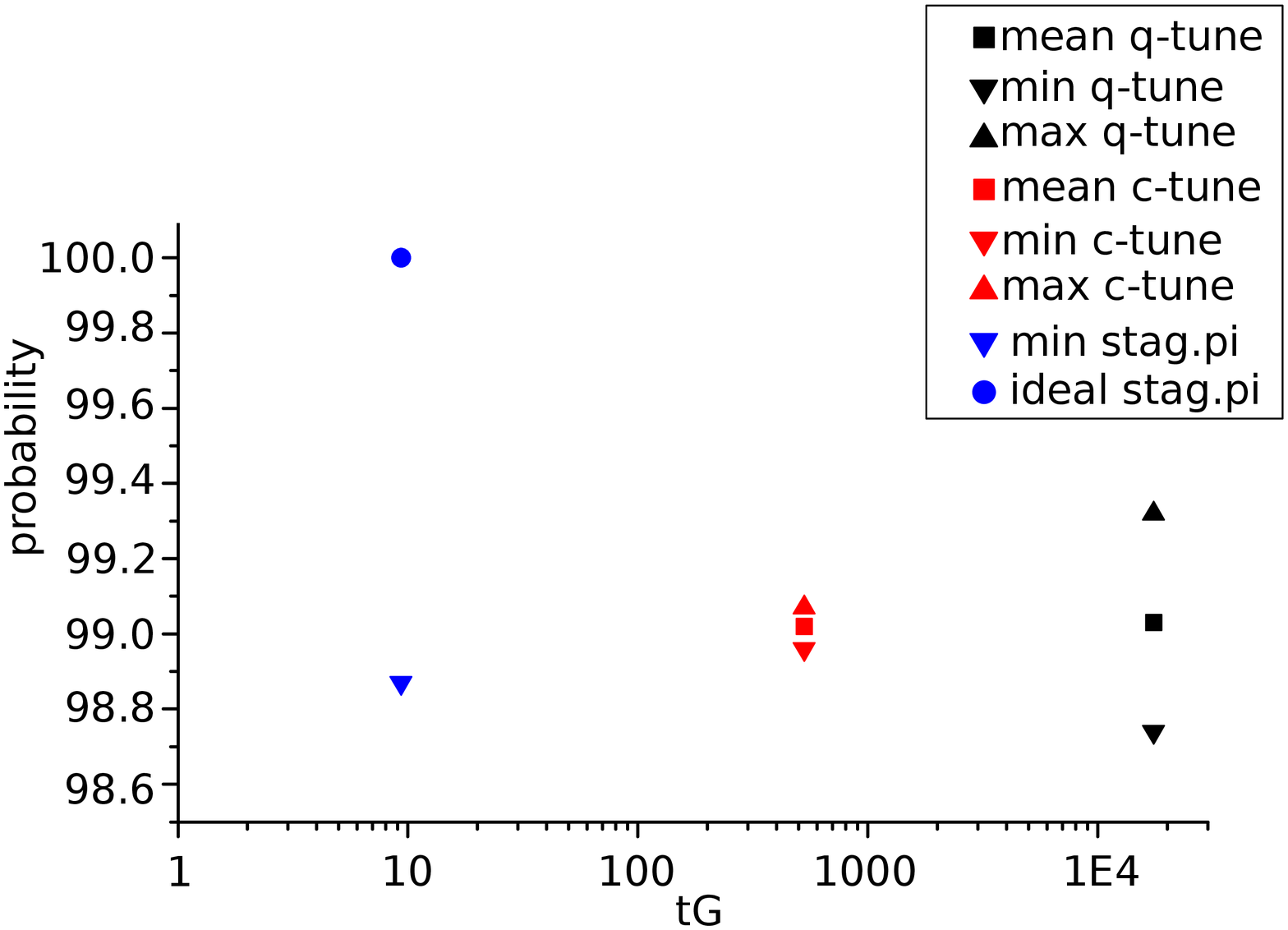}}
\caption{(color online)  Comparison of the three protocols: adiabatic sweep of the qubit frequency ("q-tune", black points), adiabatic sweep of the cavity frequency ("c-tune", red points), and resonant transfer ("stag-pi", blue points). The maximal and minimal values of the fidelity for the protocols based on adiabatic transfer  are plotted as a function of the transfer time required for achieving an average fidelity equal to 0.99. For the adiabatic protocols the finite variance is due to oscillations in the qubit population for the chosen parameters, for the resonant transfer it is due to an uncertainty of 1\% in the determination of the pulse area. In (a) $\sqrt{N}\kappa=G$, in (b) $\sqrt{N}\kappa=0.2G$. Further parameters are given in the table.\label{fig:effvglb1g1-v2_2}}
\end{figure}

These results show that transfer based on $\pi$ pulses is generally faster than adiabatic protocols. We compare the transfer time required by each type of protocol analyzed so far under the requirement that a final fidelity $\mathcal F_{\rm mw}\ge 0.99$ is achieved. In Fig. \ref{fig:effvglb1g1-v2_2}(a) we take $\sqrt{N}\kappa=G$, while in Fig. \ref{fig:effvglb1g1-v2_2}(b) the coupling strength is $\sqrt{N}\kappa=0.2G$  (with $G=2\pi\times 50$ MHz). Here, it is evident that the protocol based on resonant pulses is  two orders of magnitude faster than the protocol based on adiabatically sweeping the cavity detuning, and four orders of magnitude faster than the protocol based on sweeping the qubit detuning. The multiple data points for each protocol represent the maximal, minimal, and mean fidelity. This is extracted in different ways depending on the considered protocol. For the protocol based on adiabatic sweeping the frequencies, the population of the qubit exhibits oscillations at the asymptotics, due to the fact that the initial state is not a perfect eigenstate of the instantaneous Hamiltonian. Therefore, the mean value is the average value, while the maximal and minimal values correspond to the corresponding maximum and minimum of the oscillation.  For the resonant pulse, on the other hand, the final fidelity is affected by the pulse area, which is here taken into account by introducing an uncertainty in the pulse duration by 1\%. 

\begin{table}[tb]
  \centering
\begin{tabular}{|c||c|c|c|c|c|c|c|c|}\hline
protocol & $\sqrt{N}\kappa$ & $\delta_{Q}$& $\delta_{qk}$& $\Delta_{c}$ &$\Delta_{ck}$ \\
 \hline \hline
stag-pi & G, 0.2G &  50G & 0 & stepwise & 50G\\
c-tune & G &  7.4G & 0 & linear sweep &38G \\
            & 0.2G & 7.4G & 0 & linear sweep &26G\\
q-tune & G  & linear sweep & 4G & 50G & 0 \\
            & 0.2G & linear sweep & 0.4G & 50G & 0\\
\hline
\end{tabular}
  \caption{Parameters and their variation in time for the data points of Fig. \ref{fig:effvglb1g1-v2_2}. In the resonant pulse ("stag-pi"), in the first pulse $\delta_Q=50G$ and $\Delta_c=0$. After the excitation has been transferred to the cavity, the two values are swapped. In the adiabatic protocols, the initial value of the detuning is usually taken to be at $-\Delta_{ck}/2$ ($\delta_{qk}/2$). } 
  \label{tab:class}
\end{table}

\subsection{Effect of inhomogeneous broadening}
\label{Sec:NV}

\begin{figure}[ht!]
\includegraphics[width=7cm]{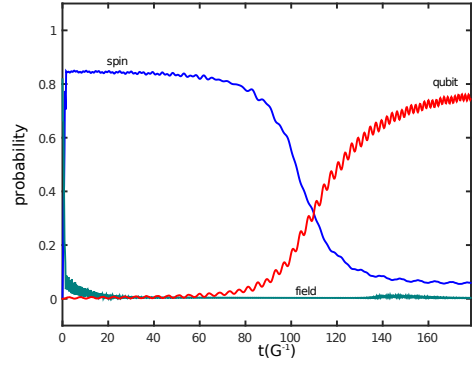}
\caption{(color online) Time evolution of the probability of spin and qubit excitation for a protocol combining EIT storage and adiabatic sweeping of the cavity frequency. We also report the excitation of the optical field as a function of time (see text). The parameters are $\Delta_{ck}=10G$, $\delta_{Q}=1.1G$. The total transfer duration is about $T_{\rm tot}=568$ns, while EIT storage is performed over a time $T_{\rm EIT}\approx g_{ab}^{-1}=$5.68ns. The curves are determined by numerically integrating Eqs. \eqref{eq:diffeqNV-final}. The inhomogeneous broadening was simulated by distributing the spins in 300 frequency values and the spectral width of the incident photon assumed to be $\Delta\omega=0.028G$.
\label{fig:KombiLauf4}}
\end{figure}

We now analyze how the fidelity $\mathcal{F}$ of the whole transfer protocol, including EIT storage, is affected by inhomogeneous broadening. For this purpose we take a medium composed by NV centers in diamond, and  assume inhomogeneous broadening of the $\ke{s}$ state of the order of 6MHz \cite{Nobauer2013} and of the order of $10$MHz of the $\ke{a}$ state  \cite{Santori2010}. The incident photon has hyperbolic secant shape. The control pulse necessary for EIT storage is identified by solving Eq. \eqref{eq:costheta}, see for instance Ref. \cite{Fleischhauer2000b}. The other parameters are: the coupling strength between optical transition and ring resonator is  $g_{ab}=2\pi \times28$MHz \cite{Barclay2009}, the decay of the upper level of the NV center $\gamma_a = 2\pi \times0.5$MHz \cite{Barclay2009}, the microwave-collective spin coupling strength is taken to be $\kappa \sqrt{N}=2\pi \times17$MHz \cite{Putz2014}, and the microwave-SCQ coupling strength is $G=2\pi \times50$MHz  \cite{Sillanpaa2007, Blais2004}, while the linewidth of the ring resonator is fixed to the value $\gamma_{co}= 2\pi \times140$MHz, which is consistent with the value of existing resonators \cite{Barclay2009,Santori2010}. Figure \ref{fig:KombiLauf4} displays the dynamics of spin and qubit excitations as a function of time for a protocol combining EIT storage with adiabatic sweeping of the cavity detuning. We also report the excitation of the optical field, which comprises the field modes outside the resonator and the ring-cavity modes. We evaluate a total fidelity of about 0.75, while the transfer protocol is performed over a time of the order of $T_{\rm tot}=568ns$.  

\begin{figure}[ht!]
\includegraphics[width=7cm]{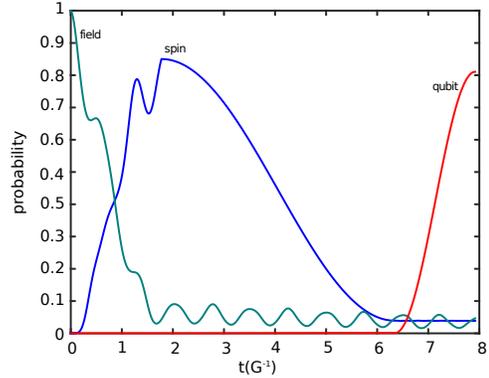}
\caption{(color online) Same as in Fig. \ref{fig:KombiLauf4} but for a protocol combining EIT storage and resonant transfer of the spin excitation to the qubit excitation. The parameters are $\delta_{Q}=28G$ and $\Delta_{ck}=28G$. The total transfer time is  of the order of 25ns.}
\label{fig:NVstagpiv3}
\end{figure}

Fidelities of about 0.8 are found by combining EIT storage with a staggered $\pi$ pulse on times of the order of $T_{\rm tot}$=25 ns, as shown in Fig. \ref{fig:NVstagpiv3}.
These calculations did not include the losses of microwave cavity and qubit, nor a possible optimization of the time variation of the parameter. The inhomogeneous broadening was simulated by distributing the spins in 300 frequency values. The parameters we took are not ideal, but of existing experiments. These results thus demonstrate the viability of the protocol. 

\section{Discussion and conclusions}
\label{Conclusions}

We proposed a protocol for transferring in a reversible way a single optical photon into the excitation of a superconducting qubit. This is realized by means of a transducer from optical to microwave regime constituted by a spin ensemble. Depending on the nature of the spin ensemble, different strategies are appropriate. Here  we considered an homogeneously-broadened ensemble of cold atoms or by NV centers in diamonds, which shall exhibit coupled electronic transitions either interfacing with the optical or with the microwave fields. We analyzed the fidelity of the transfer for different strategies, which employ EIT storage of the photon in the ensemble followed by either adiabatic or resonant transfer to the qubit excitation, mediated by the coupling with a microwave cavity. We searched for solutions that maximize the fidelity over times which are sufficiently short, to minimize the influence of detrimental effects, such as cavity decay, the dephasing of the spins, and decay processes of the qubit. The fidelity we find is essentially limited by EIT storage, and without  performing a systematic optimization results to be of the order of 0.8. It is essentially limited by the inhomogeneous broadening of the NV centers ensemble and by the parameter ranges allowing for the coherent transfer of the spin ensemble excitation to the SC-qubit via the resonator. For a spin ensemble composed by cold atoms, on the other hand, the fidelity could reach 0.99. 

We remark that this protocol requires one to combine optical and microwave cavities in the same setup, which can be experimentally realized on a chip as in Ref. \cite{Schuck:2013} or in a 3D cavity setup as it is discussed in ref. \cite{Probst2014b}. Nevertheless, the cavity must be well shielded from fields when the SCQ or the cavity are tuned, otherwise there will be trapped vortices in the cavity which results in a decrease of its quality factor \cite{Song2009}. Moreover, the SCQ must be shielded from stray optical photons, whose absorption will be to break up the Cooper pairs. For the implementation of the protocol a charge type SCQ like cooper pair box or transmon can be used if high coupling strengths between the qubit and the microwave cavity are needed.

The transfer protocols from the collective spin of the ensemble to the SCQ are based on the ability to tune qubit and cavity frequency in time. In the setups of Ref. \cite{Schuck:2013}  the cavity frequency can tuned by about $\sim$0.5\,GHz. Theoretically, a flux qubit can be tuned from its maximal frequency $\omega_{max}$ all the way down to a minimal frequency $\omega_{min}=0$. However,  flux fluctuations become more and more relevant at the steep slopes of the frequency function leading to shorter $T_2$ times. This of course limits the real tuning range of the qubits to a very limited region. A possibility to overcome this issue has been proposed in Ref. \cite{Strand2013}, where it is suggested to tailor the frequency spectrum of the qubit so that $\omega_{min}$ is not zero anymore and there are now an upper and a lower region where the slope goes to zero. In this way one can start in a region of a long $T_2$ time and stop in another region of stability. Alternatively, one could tune the spin transition, as done for a single NV center in Ref. \cite{TUne:NV}, even though this procedure is not simple to scale to the whole ensemble. When the transfer protocol is based on resonant pulses, on the other hand, one can avoid tuning the qubit. This latter protocol is by far the most efficient, even though it is more sensitive to parameter fluctuations.

\acknowledgements
We thank Britton Plourde (Syracuse University) and Sebastian Probst for discussions and helpful comments. This work has been supported by the BMBF (QuORep, Contract Nos. 01BQ1011 and 16BQ1011, and Qu.com) and by the German Research Foundation (DFG).

\begin{appendix}

\section{Hamiltonian in laboratory frame}
\label{App:A}

We report here the Hamiltonian in the laboratory frame and the transformation which allows us to write the Hamiltonian terms in Sec. \ref{Sec:Model}. For the spin ensemble, \begin{equation}
\label{H:spin:0}
\hat H_{\rm spin}^{(L)}=\hbar \sum_i^N \left(\omega_{ab}^{(i)}\hat{\sigma}_{ba}^{(i)\dagger}\hat{\sigma}_{ba}^{(i)}+ \omega_{sb}^{(i)}\hat{\sigma}_{bs}^{(i)\dagger}\hat{\sigma}_{bs}^{(i)}\right)\,,
\end{equation}
where the frequency $\omega_{ab}^{(i)}$ ($\omega_{sb}^{(i)}$) of the transition $\ke{b}\to\ke{a}$ ($\ke{b}\to\ke{s}$) in general depends on the position $\vec{r}_i$ of the spin $i$ within the sample. The Hamiltonian for the coupling with the optical fields reads
\begin{eqnarray}
\hat{H}_{\rm opt}^{(L)}
&=&\hbar \omega_{co}\left( \hat{c}_R^\dag\hat{c}_R +\hat{c}_L^\dag\hat{c}_L\right)\\
&  &+\hbar\sum_{i=1}^N\left[g_{ab}\left(\hat{c}_Re^{i\vec{k}\cdot \vec{r}_i}+\hat{c}_Le^{-i\vec{k}\cdot \vec{r}_i}\right)\hat{\sigma}_{ba}^{(i)\dagger}+{\rm H.c.}\right]\nonumber\\
& & +\hbar \sum_{i=1}^N\left(\Omega(t)e^{-i(\nu_dt-\vec{k}_L\cdot \vec{r}_i)}\hat{\sigma}^{(i)\dagger}_{sa}+{\rm H.c.}\right)\nonumber\,.
\end{eqnarray}
Here, $\hat{c}_R$ and $\hat{c}_L$ annihilate a photon of the cavity modes at frequency $\omega_{co}$ which propagates in clock- and anticlockwise direction, respectively, $\pm\vec{k}$ are the corresponding wave vectors, and  $g_{ab}$ is the vacuum Rabi frequency. The laser field is at carrier frequency $\nu_d$ couples with the transition  $\ke{s}\to\ke{a}$ with spatially-homogeneous strength $\Omega(t)$.

The controlled transfer between the spin excitation and the superconducting qubit at transition frequency $\omega_Q$  is described by the Hamiltonian operator:
\begin{eqnarray}
\label{H:mw:0}
\hat{H}_{\rm mw}^{(L)}&=&\hbar \omega_Q \hat{\sigma}^\dagger_Q\hat{\sigma}_Q+\hbar \omega_{c \mu}\hat{a}^\dag\hat{a}\\
& &+\hbar \left[\hat{a}^\dag\left(G\hat\sigma_Q+\sum_{i}^{N}\kappa_i\hat{\sigma}_{bs}^{(i)}\right)+{\rm H.c.}\right]\nn\,,
\end{eqnarray}
where $\omega_{c \mu}$ denotes the frequency of the microwave resonator.

Finally, the Hamiltonian for the dynamics of the coupling between the cavity modes and the external quantum electromagnetic field outside the resonator, takes the form
\begin{eqnarray}
\hat H_{\rm in-out}^{(L)}&=&\sum_l\sum_{p=\pm} \hbar\omega_l \hat{d}_l^{(p)\dag}\hat{d}_l^{(p)}\\
& &+\hbar \sum_l \kappa_{{\rm opt},l}\left(\hat{c}_R^\dag\hat{d}_l^{(+)} +\hat{c}_L^\dag\hat{d}_l^{(-)} +{\rm H.c.}\right)\nn\,,
\end{eqnarray}
where the modes have frequencies $\omega_l$, wave vector $k_l$ ($-k_l$) along the positive (negative) direction in the $x$ axis, and corresponding annihilation and creation operators $\hat{d}_l^{(+)}$ and $\hat{d}_l^{(+)\dag}$ ($\hat{d}_l^{(-)}$ and $\hat{d}_l^{(-)\dag}$), respectively.

It is convenient to move to the reference frame defined by the unitary transformation $\hat U=\hat U_0\otimes \hat U_1\otimes \hat U_2\otimes\hat U_3$, where the individual unitary transformations of the tensor product commute and take the form  
\begin{eqnarray*}
&&\hat U_0=\exp\left[it\omega_{co}(\hat{c}_R^\dag\hat{c}_R +\hat{c}_L^\dag\hat{c}_L-\sum_i\ke{b}_i\br{b})\right]\,,\\
&&\hat U_1=\exp\left[-it\nu_d\sum_i\ke{s}_i\br{s}\right]\,,\\
&&\hat U_2=\exp\left[it\Delta(\ke{e}\br{e}+\hat a^\dagger \hat  a)\right]\,,
\end{eqnarray*}
with $\Delta=\omega_{co}-\nu_d$ the detuning between cavity and laser fields coupling the states $\ke{b}\to\ke{s}$ via Raman scattering through state $\ke{a}$, 
while $\hat U_3=\exp\left[it\omega_{co}\sum_{l,p}\hat{d}_l^{(p)\dag}\hat{d}_l^{(p)}\right]$. The transformed Hamiltonian is $\hat H=\hat U \hat H^{(L)}\hat U^\dagger-i\hbar\hat U\partial_t\hat U^\dagger$ and the explicit time dependence of the classical fields disappears.  In this reference frame, the other relevant frequency shifts are the detunings with respect to the reference fields of the external modes, $\Delta_l=\omega_l-\omega_{co}$, of the optical transition $\delta_{ab}^{(i)}=\omega_{ab}^{(i)}-\omega_{co}$, of the magnetic dipole transition $\delta_{sb}^{(i)}=\omega_{sb}^{(i)}-\Delta$, of the microwave cavity $\Delta_{c}=\omega_{c\mu}-\Delta$, and of the qubit transition $\delta_{Q}=\omega_Q-\Delta$. 

\section{Basic equations}
\label{App:equations}
The equations of motion of the probability amplitudes take the form
 \begin{subequations}
  \label{eq:diffeqNV-final}
   \begin{eqnarray}
   \label{eta+}
\dot{\eta}_l^{(+)}(t)&=&-i\Delta_l \eta_l^{(+)}(t)-i\kappa_{{\rm opt},l} u(t)\,,\\
\label{eta-}
\dot{\eta}_l^{(-)}(t)&=&-i\Delta_l \eta_l^{(-)}(t)-i\kappa_{{\rm opt},l} v(t)\,,\\
\dot{u}(t)&=&-i\sum_i^N g_{ab}a_i(t)e^{-i\vec{k}\cdot \vec{r}_i}-i\sum_l\kappa_{{\rm opt},l} \eta_l^{(+)}(t)\nn\\
&\quad &-\frac{\gamma_{co}}{2}u(t)\,,\\
 \dot{a}_i(t)&=&-\left(i\delta_{ab}^{(i)}+\frac{\gamma_a}{2}\right)a_i(t)-i\Omega(t)s_i(t)e^{i\vec{k}_L\cdot \vec{r}_i}\nn\\
 & &-ig_{ab}\left(u(t)e^{i\vec{k}\cdot \vec{r}_i}+v(t)e^{-i\vec{k}\cdot \vec{r}_i}\right)\,,\\
 \dot{v}(t)&=&-i\sum_i^N g_{ab}a_i(t)e^{+i\vec{k}\cdot \vec{r}_i}-i\sum_l\kappa_{{\rm opt},l} \eta_l^{(-)}(t)\nn\\
&\quad &-\frac{\gamma_{co}}{2}v(t)\,,\\
 \dot{s}_i(t)&=&-\left(i\delta_{sb}^{(i)}+\frac{\gamma_s}{2}\right)s_i(t)-i\Omega(t)^{*}e^{-i\vec{k}_L\cdot\vec{r}_i}a_i(t)\nn\\
 & &-i\kappa_i c(t)\,,\\
 \dot{c}(t)&=&-\left(i\Delta_{c}(t)+\frac{\gamma_{c\mu}}{2}\right)c(t)-i \sum_i^N\kappa_is_i(t)\nn\\
&\quad &-iGq(t)\,,\\
 \dot{q}(t)&= &-i\left(\delta_{Q}q(t)+\frac{\gamma_e}{2}\right)q(t)-iGc(t)\,,
  \end{eqnarray}
 \end{subequations}
where $\gamma_{\ell}$ is the decay of the $\ke{\ell}$ state, with $\ell=a,s,e$, while  $\gamma_{co}$ and $\gamma_{c\mu}$ are the decay rates of optical ring and microwave cavity modes.  The norm of the state vector is thus not conserved assuming that irreversible processes can couple the considered set of states to other states. 

\section{EIT storage}
\label{App:EIT}

We review the basic steps in order to warrant that a single incident photon is adiabatically transferred to a spin excitation of the medium, closely following the treatment in Ref. \cite{Fleischhauer2000b}. Our goal is to maximize the efficiency of transferring the initial state, Eq. \eqref{state:init}, into the target intermediate state $$\ke{\Psi}_{\rm opt | target}=\ke{vac}\ke{0_L,0_R}\ke{s}_e\,,$$ with $\ke{s}_e=\sum_j\ke{s}_j/\sqrt{N}$ the symmetric Dicke state. For this purpose we first assume that the spin ensemble is not coupled to the microwave cavity field, which can be achieved by setting the cavity mode and the qubit out of resonance. We assume a homogeneously broadened atomic ensemble, neglect decay of the metastable states, corresponding to setting $\gamma_s=0$ in Eqs. \eqref{eq:diffeqNV-final}. The laser propagation direction is parallel to the one of the incident photon, so that $\vec{k}_L=\vec{k}$. Furthermore, we assume that the frequency width of the incident photon is much smaller than the cavity linewidth, which allows us to take $\kappa_{{\rm opt},l}\simeq \kappa_{\rm opt}$. Under these assumptions it is convenient to introduce $a(t)=\sum_je^{-i\vec{k}\cdot \vec{r}_j}a_j(t)/\sqrt{N}$ and  $s(t)=\sum_js_j(t)/\sqrt{N}$, which are the amplitudes of the collective electronic excitations due to the coupling with the clockwise cavity mode and the laser. The equations of motion can be then cast in the form:
\begin{subequations}
  \label{eq:diffeq:2}
   \begin{eqnarray}
\dot{u}(t)&=&-i\sqrt{N}g_{ab}a(t)-i\kappa_{\rm opt}\sum_l \eta_l^{(+)}\nn\\
& &-\frac{\gamma_{co}}{2}u(t)\,,\\
 \dot{s}(t)&=&-i\Omega(t)^{*}a(t) \,,\\ 
\dot{a}(t)&=&-\frac{\gamma_a}{2}a(t)-i[\Omega(t)s(t)+\sqrt{N}g_{ab}u(t))]\nn\\
 & &-ig_{ab}\sqrt{N}v(t)\left(\sum_ie^{-2i\vec{k}\cdot \vec{r}_i}/N\right)\,,\\
 \dot{v}(t)&=&-i g_{ab}\sqrt{N}\left(\sum_i^Na_i(t)e^{+2i\vec{k}\cdot \vec{r}_i}/N\right)\nn\\
 & & -i\sum_l\kappa_{{\rm opt},l} \eta_l^{(-)}(t)
-\frac{\gamma_{co}}{2}v(t)\,,
 \end{eqnarray}
 \end{subequations}
 where the equations for the amplitudes of the incoming and outcoming photons are given by Eqs. \eqref{eta+}-\eqref{eta-}.
Insight can be gained by moving to the basis of the dark and bright states $\ke{D}$ and $\ke{B}$ for the internal and optical cavity excitations, defined as \cite{Fleischhauer2002}
\begin{align}
\label{eq:dark-bright}
\ke{D}&=-{\rm e}^{-{\rm i}\phi}\cos \Theta(t)\ke{b}_e\ke{1_R}+\sin \Theta(t)\ke{s}_e\ke{0_R}\,,\\
\ke{B}&=\sin \Theta(t)\ke{b}_e\ke{1_R}+{\rm e}^{{\rm i}\phi}\cos \Theta(t)\ke{s}_e\ke{0_R}\,,
\end{align}
where $\tan \Theta(t)=g_{ab}\sqrt{N}/\Omega_0(t)$. The corresponding probability amplitudes take the form $D(t)=-{\rm e}^{-{\rm i}\phi}\cos \Theta(t)u(t)+\sin \Theta(t)s(t)$ and $B(t)=\sin \Theta(t)u(t)+{\rm e}^{{\rm i}\phi}\cos \Theta(t)s(t)$. The dynamics can be reduced to coupling the dark state with the incident photon for $|\tan\Theta|\ll 1$ and assuming that the time scale over which $\Theta$ varies is sufficiently long, namely, $|\dot\Theta|\ll\gamma_a$. In addition, the coupling with the other mode of the ring resonator can be neglected when $|\langle\sum_i^Na_i(t)e^{+2i\vec{k}_{co}\cdot \vec{r}_i} \rangle|\ll N$, which is satisfied for a sufficiently disordered medium. In this limit Eqs. \eqref{eq:diffeq:2}, together with Eqs. \eqref{eta+}-\eqref{eta-}, can be reduced to the coupled equations \cite{Fleischhauer2000b}:
\begin{subequations}
\label{eq:diffgl3}
\begin{align}
\dot{\eta}_l(t)&=-i\Delta_l\eta_l(t)-i\kappa_{\rm opt}\cos \Theta(t)D(t)\,, \label{eq:diffgl3-2}\\
\dot{D}(t)&=-i\kappa_{\rm opt}\cos \Theta(t)\sum_l\eta_l(t)\nonumber\\
&\simeq  -i\sqrt{\gamma_{co} \frac{c}{L}}\cos \Theta(t) \Phi_{in}(0,t)-\frac{\gamma_{co}}{2}\cos ^2 \Theta (t)D(t)\,,\label{eq:Ddot2}
\end{align}
\end{subequations}
where the coupling to the other states is negligible in the adiabatic limit and we applied the Markov approximation.  In writing Eq. \eqref{eq:Ddot2} we have denoted by $\Phi_{in}(0,t)=\sum_l{\rm e}^{-i\Delta_lt}\eta_l^{(+)}(0)$ the envelope of the input photon at the cavity mirror, $z=0$, and introduced $\gamma_{co}=\kappa_{\rm opt}^2L/c$ the cavity decay rate, with $L$ the cavity length. Perfect transfer of the photonic excitation into a spin excitation of the medium is reached by requiring that each component $\eta_l^{(+)}(t)$ vanishes, such that the output field $\Phi_{out}(0,t)=\sum_l\eta_l^{(+)}(t)$, namely, the field at $z=0$ and at the instants of time in which the excitation shall be absorbed, is zero. The output field is found by solving Eqs. \eqref{eq:diffgl3}, assuming that $D(t)=0$ for all times $t \le 0$, and reads (after setting $D(t)\to iD(t)$) \cite{Fleischhauer2000b}
\begin{align}
\label{eq:Phi1}
\Phi_{out}(0,t)&=\Phi_{in}(0,t)-\sqrt{\gamma_{co} \frac{L}{c}}\cos \Theta(t) D(t)\,,
\end{align}
with
\begin{align}
\label{eq:D1}
D(t)&=\sqrt{\gamma_{co} \frac{c}{L}}\int_{t_0}^t d\tau \cos \Theta(\tau)\Phi_{in}(0,\tau)\nonumber\\
&\quad \times \exp\left(-\frac{\gamma_{co}}{2}\int_{\tau}^{t} d\tau '\cos ^2\Theta(\tau ')\right)\,.
\end{align}
Impedance matching, here corresponding to the destructive interference of the directly reflected and the circulating components, requires  $\Phi_{out}=\dot{\Phi}_{out}=0$, which 
is found when the fields satisfy the equation: 
\begin{align}
\label{eq:input-output}
-\frac{d}{dt}\ln \cos \Theta (t)+\frac{d}{dt} \ln \Phi_{in}(t)&=\frac{\gamma_{co}}{2}\cos^2 \Theta (t)\,.
\end{align}
For an incident photon envelope given by a normalized hyperbolic secant form, 
\begin{align}
\label{eq:inputpulse}
\Phi_{in}(z=0,t)&=\sqrt{\frac{L}{cT}}{\rm sech} \left[\frac{2t}{T}\right]\,,
\end{align}
then Eq. \eqref{eq:input-output} can be simply solved (assuming the asymptotic behavior $\cos \Theta \rightarrow 0$ for $t \rightarrow \infty$) by 
\begin{align}
\cos \Theta(t) &=\sqrt{\frac{2}{\gamma_{co} T}}\frac{{\rm sech}(2t/T)}{\sqrt{1+ \tanh \left[2t/T\right]}}\,.
\label{eq:costheta}
\end{align} 
This formula delivers the ideal time shape of the pump-laser field $\Omega_0(t)$ which warrants a perfect transfer of the single incident photon to an ensemble spin excitation. 

\end{appendix}

\end{document}